\documentclass[sigconf,screen]{acmart}

\AtBeginDocument{%
  }

\copyrightyear{2026}
\acmYear{2026}
\setcopyright{cc}
\setcctype{by-nc-nd}
\acmConference[AutomotiveUI '26]{18th International Conference on Automotive User Interfaces and Interactive Vehicular Applications}{September 20--23, 2026}{Gothenburg, Sweden}
\acmBooktitle{18th International Conference on Automotive User Interfaces and Interactive Vehicular Applications (AutomotiveUI '26), September 20--23, 2026, Gothenburg, Sweden}
\acmDOI{10.1145/3828157.3828775}
\acmISBN{979-8-4007-2814-3/2026/09}

\PassOptionsToPackage{table}{xcolor}

\usepackage{pifont}
\newcommand{\xmark}{\color{red}\ding{55}}%
\usepackage{pifont}
\usepackage{array}
\usepackage{xcolor}
\usepackage{colortbl}
\usepackage{longtable}
\begin{document}

\title[Inconsistent by Design]{Inconsistent by Design: A Systematic Review of Experimental Design Practices Across Impaired Driving Domains}


\author{Kayli Battel}
\email{kayli.battel.ctr@tri.global}
\orcid{0009-0003-2300-8718}
\affiliation{%
  \institution{Toyota Research Institute}
  \city{Cambridge}
  \state{Massachusetts}
  \country{USA}
}

\author{John Gideon}
\email{john.gideon@tri.global}
\orcid{0000-0003-3945-3341}
\affiliation{%
  \institution{Toyota Research Institute}
  \city{Cambridge}
  \state{Massachusetts}
  \country{USA}
}

\author{Megan Applegate-Kenton}
\email{megan.applegate-kenton@tri.global}
\orcid{0000-0001-8111-7312}
\affiliation{%
  \institution{Toyota Research Institute}
  \city{Cambridge}
  \state{Massachusetts}
  \country{USA}
}

\author{Patricio Reyes Gomez}
\email{patricio.reyesgomez.ctr@tri.global}
\orcid{0009-0003-7078-2044}
\affiliation{%
  \institution{Toyota Research Institute}
  \city{Cambridge}
  \state{Massachusetts}
  \country{USA}
}

\author{Anshul Gupta}
\email{anshul.gupta@tri.global}
\orcid{0009-0004-0388-6350}
\affiliation{%
  \institution{Toyota Research Institute}
  \city{Cambridge}
  \state{Massachusetts}
  \country{USA}
}

\author{Todd Rowell}
\email{todd.rowell@tri.global}
\orcid{0009-0009-0060-4465}
\affiliation{%
  \institution{Toyota Research Institute}
  \city{Cambridge}
  \state{Massachusetts}
  \country{USA}
}

\author{Thomas M. Balch}
\email{thomas.balch@tri.global}
\orcid{0009-0002-5112-7178}
\affiliation{%
  \institution{Toyota Research Institute}
  \city{Cambridge}
  \state{Massachusetts}
  \country{USA}
}

\author{Emily Sarah Sumner}
\email{emily.sumner@tri.global}
\orcid{0000-0003-1912-9640}
\affiliation{%
  \institution{Toyota Research Institute}
  \city{Cambridge}
  \state{Massachusetts}
  \country{USA}
}

\author{Guy Rosman}
\email{guy.rosman@tri.global}
\orcid{0000-0002-9334-1706}
\affiliation{%
  \institution{Toyota Research Institute}
  \city{Cambridge}
  \state{Massachusetts}
  \country{USA}
}

\begin{abstract}
    Impaired driving, including distraction, fatigue, and intoxication, leads to thousands of fatalities annually. Impairment detection and related assistive technologies are rapidly advancing, but their full potential remains unrealized. The inconsistent maturity of impairment research and anomalies within individual domains are key barriers, including variations in taxonomic characterization, experimental data-collection methodologies, and treatment of impairments across studies.

We present a unified review of impairment studies, covering 91 studies across nine impairment domains coded on three dimensions: impairment induction methods, scenario hazards, and observable phenomena and metrics. Our key findings include: (1) a proposed performance-degradation vs. event-response paradigm for existing impairment domains, (2) a maturity framework along the three dimensions, and (3) an analysis of impairment research methodologies, revealing anomalies and gaps in the treatment of scenarios and metrics. We further propose recommendations for standardizing methodologies to support future cross-domain research and development of holistic detection and assistive systems.
\end{abstract}
\renewcommand{\shortauthors}{Battel et al.}


\maketitle
\let\thefootnote\relax\footnotetext{This is the authors' version of the work. It is posted here for personal use. Not for redistribution. The definitive Version of Record will be published in AutomotiveUI '26, \url{https://doi.org/10.1145/3828157.3828775}.}
\section{Introduction}
\label{sec:Introduction}
Driver impairments such as distraction, alcohol intoxication, and fatigue lead to thousands of accidents and fatalities each year~\cite{NCSA2025distracted,nhtsa2017drowsy,nhtsa2024alcoholimpaired,nhtsa2023economic}. Naturalistic studies have reported that impairments and driver state factors, including distraction, emotional distress, and other risky factors contribute to the majority of recorded crashes ~\cite{dingus2016driver}. However, studies have consistently demonstrated that a convergence of factors involving driver state, as well as vehicle and scene factors contributes to the conditions that result in an accident ~\cite{singh2015critical}, emphasizing the need for research that examines impaired driving across diverse vehicle and scene conditions and interactions with safety-critical events to meaningfully inform assistive technology development. While technology to detect and assist in cases of impaired driving is advancing~\cite{Kashevnik2021_59, prendez2024assessment,ayas2024drowsiness,gideon2025simulator}, a deeper understanding of how impairments interact with driver behavior, driving risk, and assistive systems — and with one another — remains necessary to improve intervention design. While \citeauthor{seaman2025identifying} ~\cite{seaman2025identifying} argue that an integrated detection and intervention approach should be pursued based on shared physiological and behavioral markers across driver states, our review asks the prior question: whether the methodological infrastructure currently exists to support that integration. Developing the required understanding necessitates controlled experimental data that is difficult to collect on public roads (e.g., exposing impaired drivers to collision-imminent scenarios or sudden obstacles); human-in-the-loop simulator studies remain the primary source for such data~\cite{tejero2006concept,Kang2017_9}, and merit systematic investigation of the state of the art across impairment domains.

Driver impairments are often studied in isolation, with inconsistent methods across domains. Meaningful cross-domain synthesis requires more than aggregating domain-specific findings; it requires a shared analytical framework. Prior reviews have examined individual or paired impairment domains, but none have applied a unified coding scheme across structurally dissimilar domains to determine whether their methods are comparable. By doing so, our structural comparison reveals patterns invisible to domain-specific reviews, such as the isolation of performance-degradation paradigms (study designs that continuously evaluate metrics during impairment, e.g., ~\cite{strayer2003cell}) and event-response paradigms (study designs that evaluate metrics in response to a hazard, e.g., ~\cite{greenberg2003distraction}) within domains, which domain-specific reviews cannot reveal.

Domain-specific reviews exist for fatigue~\cite{sikander2018driver}, distraction~\cite{Young2007,ge2022review}, and intoxication~\cite{garrisson2021effects,buckley2026rapid} individually, with some two-domain surveys (for example, distraction and intoxication~\cite{Shiferaw2014_32,ahlstrom2023alcohol} or fatigue and intoxication~\cite{dingus1987development}), and yet cross-domain structural differences appear only when coding them simultaneously within the same framework. No existing review has characterized these domains along key shared methodological dimensions — \textit{induction method} (the experimental procedure used to cause or elicit a target impairment state in a driver), \textit{scenario design} (the simulator or on-road conditions, setup, and elements), and \textit{hazard structure} (a stimulus or safety-critical event embedded in a driving scenario that demands a driver response)~\cite{sagberg2015review} — surfacing the need for more integrated frameworks. 

This review addresses this gap to inform our own study design approach and contribute to the field of driver impairment research by surveying human-subject studies on driving impairments, mapping out key aspects (e.g. induction methods, properties of scene and driving hazards~\cite{sagberg2015review}, measures of observable phenomena), and their relation to assistive interactions. For the purpose of the survey, we identified and explored nine common impairments that appeared in the literature: (1) cognitive load~\cite{engstrom2017effects}, (2) cognitive distraction~\cite{strayer2011cognitive}, (3) visual distraction~\cite{strayer2011cognitive}, (4) alcohol intoxication~\cite{ogden2004effects}, (5) fatigue~\cite{brown1994driver}, ~\cite{horne1995sleep}, (6) frustration~\cite{shinar1998aggressive}, (7) stress~\cite{matthews1996validation}, (8) sudden sickness~\cite{dobbs2001medical}, and (9) readiness~\cite{fuller2005towards,mioch2017driver}. For definitions of each domain, see the associated citations.

The review process was informed by the following research questions (RQs), which inform the three main dimensions we assess impairment domains on throughout this review: 

\indent RQ1. How are induction methods standardized or varied within and across domains? 

\indent RQ2. Which hazard paradigms are used and how consistently within and across domains?

\indent RQ3. How do observable phenomena and metrics vary within and across domains?  

By applying a shared framework across multiple impairment domains, we identified systematic differences in how impairment domains are studied. These manifest in three main findings:  (1) evidence of a performance-degradation vs. event-response paradigm split in driver performance metrics, (2) uneven methodological maturity across the three domain dimensions, and (3) domain-specific anomalies that reveal deeper inconsistencies in how the field conceptualizes risk. Together, these findings motivate a meta-framework for impaired driving research that addresses unified taxonomies, the interplay of impairment domains, driver behavior in context of the environment, and the role of assistive systems.
\section{Background}
\label{sec:Background}

Significant research effort has been dedicated to driver impairment, an overall term for psychophysiological status that is insufficient to sustain a safe level of vehicular control~\cite{Brookhuis2003_27}. According to the National Highway Traffic Safety Administration critical reason framework, crash causation is categorized across driver, vehicle, environment, and driver-related factors, with the latter contributing the most to causation ~\cite{singh2015critical}. Driving statistics have associated different types of impairment conditions with thousands of fatalities yearly in the US ~\cite{NCSA2025overview}.

Distracted driving with various types of distraction~\cite{Young2007} represents a substantial and well-documented road safety hazard. Naturalistic studies have provided insight into the pervasiveness of distraction in every-day driving; a large-scale study of 3,542 drivers in the U.S. reported that distraction was a factor in 68\% of recorded crashes, with drivers distracted 52\% of the time, overall ~\cite{dingus2016driver}. Despite the focus on phone use as a source of distraction, other risky behaviors like reading or writing and even emotional distress significantly increased crash odds compared to non-distracted driving ~\cite{dingus2016driver}.

Human-in-the-loop simulator studies, which involve participants using a simulated system (steering wheel and pedals controlling a virtual vehicle displayed on a monitor) to virtually experience various driving scenarios while physiological and driving performance data is collected, enable controlled study of conditions that would otherwise be unsafe or unethical to induce on public roads ~\cite{mullen2011simulator,wynne2019systematic,himmels2023towards}, making them the predominant platform for impaired driving research. Moreover, machine learning (ML)-driven assistive systems necessitate data on impaired driving in nominal and near-accident scenarios, for which such simulator studies are essential~\cite{dargahi2024multimodal,gideon2025simulator}. Simulator studies are treated as valid primary evidence because research questions concern experimental design practices rather than absolute behavioral validity, and the ecological validity implications of simulator dominance are addressed in the Results section. 

While many existing reviews focus on individual types of impairment ~\cite{irwin2017alcohol,soares2020drowsiness,kabilmiharbi2022workload}, others have examined a subset of the impairment domains considered in this review. Example researched subsets included intoxication and distraction (visual and cognitive) ~\cite{Shiferaw2014_32, harrison2011alcohol},  intoxication and drowsiness ~\cite{dingus1987development}, stress and frustration~\cite{CunninghamRegan2016emotion}, and variable distraction types ~\cite{Young2007}. Notably, while these reviews cover multiple adjacent types of impairment, they do not apply a shared framework across structurally dissimilar impairment domains simultaneously.
As noted above, ~\citeauthor{seaman2025identifying} ~\cite{seaman2025identifying} make the case for integrated detection and intervention; our review provides the methodological audit that such integration would require. The three dimensions (impairment induction, hazards, and metrics) reflect a common structure of impaired driving research: researchers induce an impairment state, embed safety-critical events to elicit driver responses, and measure the resulting behavior~\cite{Brookhuis2003_27}. By systematically coding studies along three shared dimensions, we find that the current methodological infrastructure underlying each domain is too heterogeneous to support valid cross-domain comparison. 

Traditional approaches for impairment detection focused on individual categories, such as visual distraction or cognitive distraction~\cite{liang2007real,MedeirosWard2015_53,misra2023detection}. However, in recent years, approaches have started to look at the detection of tuples of distraction types~\cite{gideon2025simulator,abbas2022hypo,ahlstrom2023alcohol}, or multiple sub-types of manual distraction~\cite{yang2023quantitative,ma2024vit}. These recent results merit a more consistent treatment of the data collected in studies with multiple impairment domains, and better uniformity in our exploration of driver impairment in general.

Finally, principled data curation has been shown to be critical for ML performance across a range of fields~\cite{raffel2020exploring,fang2023data,eckhoff2023sages,grattafiori2024llama,agia2025cupid}; a shared methodological framework for impaired driving studies would similarly support the construction of datasets suitable for training and evaluating detection and assistive systems.

\section{Methods}
This review was conducted based on the Preferred Reporting Items for Systematic Reviews and Meta-Analyses (PRISMA, ~\cite{page2021prisma}) guidelines to characterize how impaired-driving research is designed across a broad range of impairment domains. In the survey, we were primarily interested in the methodological choices studies made: which impairment induction methods were used, which hazards or critical events were built into the scenarios, and which observable phenomena and metrics were measured to assess impairment. This is reflected both in the search strategy, as well as the organization of the results along the key RQs.

\subsection{Search Strategy}
\begin{figure}
    \includegraphics[width=0.9\linewidth]{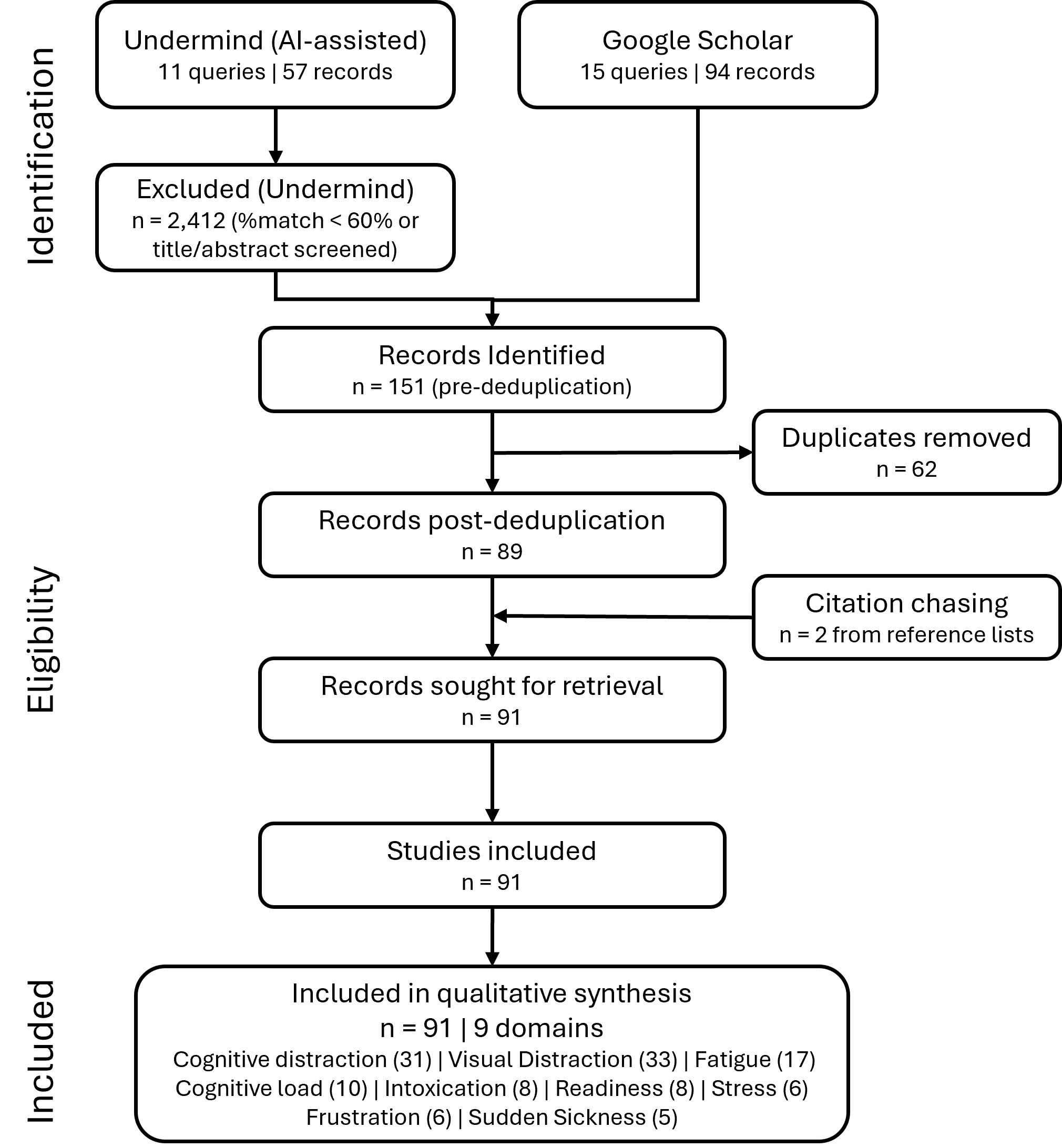}
    \caption{{PRISMA flowchart with the article selection process.}}
    \Description{The search strategy began with identification of candidate literature via Undermind (AI-Assisted) and Google Scholar searches. Eligibility was then confirmed for each candidate and the qualitative synthesis ultimately included 91 articles across nine domains.}
    \label{fig: PRISMA flowchart}
\end{figure}
To identify relevant literature, we performed Google Scholar ~\cite{googlescholar} searches and supplemented these with the AI-assisted search tool, Undermind ~\cite{hartke2024undermind}, which extended coverage across impairment domains for which standard keyword searches returned limited or incomplete results. The full set of Google Scholar search terms is reported in Table \ref{tab:S1_google_scholar} in the supplementary material.

Google search terms which produced irrelevant or redundant results included: Driving and stress, Driving and stress and performance, Stress and ``driving performance'', Frustration and vehicle and driving and performance, Driving and handoff and (surprise or readiness), ``Automated driving'' and (handoff or handover) and (surprise or readiness).

The full set of Undermind search queries is reported in Table \ref{tab:S2_undermind} in the supplementary material. This review was designed for breadth across impairment domains, not exhaustive coverage of any single domain. Lower article counts for some topics reflect both search difficulty and deliberate scoping decisions, rather than incomplete review.

\subsection{Eligibility Criteria}
We included simulator and on-road studies of driver impairment that reported induction methods, behavioral metrics, and/or scenario designs featuring structured hazard events, with primary focus on simulator studies. Initial goals for the survey were to identify approximately 90 articles from the 9 impairment domains, with about 10 articles from each domain. However, it became apparent quickly that certain domains had uneven overlap with other domains, resulting in a varied distribution.

Manual distraction ~\cite{cdc_distracted_driving} was excluded because it describes a physical behavior rather than an internal driver state, and its detection relies on a methodology distinct from the impairment domains we highlighted in the introduction -- which are inferred from physiological, ocular, or vehicle-kinematic signals. Manual distraction detection employs approaches closer to pose and activity recognition, with the benchmarking methodology often used in today's data-driven computer vision~\cite{cao2019openpose,statefarm2016}, placing it closer to computer vision methodologies~\cite{li2020detection,
yuan2021owipa}.

We excluded results that were not empirical human-subject studies (reviews, opinion pieces, validation studies, detection approaches, theoretical/policy discussions), with exceptions being made for reviews if the review focused on details of the study design practices in its articles. Studies examining chronic medical conditions, focusing on the causes of impairment rather than effects on driving behavior, demographic or individual subject differences  (e.g., driver experience), or addressing adjacent behavioral constructs to those we included (e.g., road rage), were also excluded. Additionally, studies conducted in aviation or trucking contexts, population-specific studies, and publications with quality or accessibility issues were excluded. Finally, once a domain reached saturation (i.e., once we concluded a comprehensive perspective of methodologies was represented), further articles in that domain were excluded.

\subsection{Articles searched}
Table \ref{tab:Search Result Summary} presents the total number of articles reviewed for each impairment domain, in terms of both the total number of articles that investigated that impairment (including interactions), as well as the number of articles that studied that impairment in isolation. For a full reference list of all articles reviewed, see Table \ref{tab:search_result_summary} in the Supplementary Material. We followed the proposed PRISMA–Transparent Reporting of Artificial Intelligence in Comprehensive Evidence Synthesis (PRISMA-trAIce) checklist ~\cite{holst2025transparent} as a reporting guide for the article search and selection process. 

\textbf{Screening Process.} Search results were screened in two phases. In the first phase, titles and abstracts were manually reviewed for relevance to the search intent, allowing an initial manual pass to filter clearly out-of-scope work. Articles that passed this check advanced to full-text review, where a more detailed manual assessment was made against the inclusion and exclusion criteria. Article counts at each phase are summarized in the PRISMA~\cite{page2021prisma} flow diagram (Figure \ref{fig: PRISMA flowchart}).

\begin{table} [!htbp]
    \centering
    \small
    \renewcommand{\arraystretch}{1.25}
\caption{Summary Table of Impairment Article Counts}
\Description{Table showing the number of articles reviewed per impairment domain, split into total articles (which may include studies of the domain alongside co-occurring impairments) and isolated articles (which study only that domain). Ten rows cover the nine impairment domains plus an "Other" category. Total article counts range from 5 (Sudden Sickness) to 33 (Visual Distraction). The ratio of isolated to total articles varies substantially across domains: Sudden Sickness and Frustration are studied entirely in isolation, while Readiness has zero isolated studies out of eight.}
\label{tab:Search Result Summary}
    \begin{tabular}{@{}>{\centering\arraybackslash}p{0.4\linewidth}>{\centering\arraybackslash}p{0.25\linewidth}>{\centering\arraybackslash}p{0.25\linewidth}@{}}\toprule
         Topic&  Total& Isolated\\\midrule
Cognitive Load&  
10& 
3\\
Cognitive Distraction&  
31& 
11\\
Fatigue& 
18& 
10\\
Readiness&
8&
0\\
Alcohol 
Intoxication&
8&
3\\
Sudden Sickness&
5&
5\\
Stress&
6&
4\\
Frustration&
6&
6\\
Visual Distraction&
33&
10\\
Other&
12&
4\\ \bottomrule
    \end{tabular}
\end{table}

\textbf{Additional Impairment Categories.} Several articles in the review incorporated ``other'' impairments with the following topics: obstructive sleep apnea (1); drowsiness as a result of automation (1); manual distraction from tasks involving physical manipulation of a stimulus (5); permanent visual impairments including simulated cataracts, binocular visual field restriction, monocular condition, blurred vision (2); visual ``performance,'' meaning ability to stay on course after looking away from the road (1); impairment definition and assessment review (1); and takeover induction via various buzzer, vibration, LED, etc. alerts (1). We did not search for these terms directly, but these topics were categorized as ``other'' when they appeared in articles of interest. Note that we found two articles with permanent visual impairment before determining that this domain was not comparable or of interest and excluded such articles afterward.

\subsection{Data Extraction and Coding Framework}

\textbf{Coding Schema.} We extracted data from each included article using a structured coding schema developed to capture the methodological dimensions central to our RQs. Existing domain-specific simulator reviews apply structured coding to study design features within a target impairment. For example, ~\citeauthor{Shariati2025}~\cite{Shariati2025} code independent variables (alert type, takeover time budget, non-driving-related task), dependent variables, and simulator parameters for the takeover/readiness domain. Our primary coding sheet adopts the same scaffold for metadata, study format, and performance measures. However, we needed to generalize domain-specific fields from single-domain reviews (e.g., induction method and scenario hazard events) to accommodate structurally dissimilar impairments within the same instrument. For each article, we recorded in a Google Sheet spreadsheet: metadata (title, year, access, and search terms), sample size, study format (simulator, on-road, or combined), impairment induction method, use of driver assistance systems, route type and environmental conditions, traffic density and types, hazard or reaction-inducing events, and the measurements, sensors, and variables collected. Undermind AI results were screened first by title, abstract, and relevance-to-prompt percentage, and then by full text ~\cite{holst2025transparent}. They were then coded manually using the same process as Google Scholar results, with more information available in Section \ref{sec:S3.8_limitations_of_AI} of the Supplementary Material.

This primary coding sheet informed three follow-up instruments covering specific dimensions relevant to cross-domain study design comparison, as discussed in Section \ref{sec:Background}: induction approaches (impairment domain, induction name, protocol description), incident and hazard scenarios (hazard name, impairment domain, scenario description), and observable phenomena and metrics (induction, metric name, objective or subjective classification, canonical category, and required instrumentation).

Applying these fields uniformly across all nine domains surfaced structural dissimilarities: where single-domain reviews can use pre-specified values (e.g., a fixed set of takeover alert types for readiness~\cite{Shariati2025}), cross-domain coding requires category definitions general enough to accommodate, for example, blood alcohol content (BAC) manipulation, sleep restriction, and scenario-based frustration within the same column. Differences in how those columns populate across domains constitute the structural findings reported in Section~\ref{sec:Results}. These three instruments correspond directly to the RQs: RQ1 relates to induction approaches, RQ2 relates to incidents and hazards, RQ3 relates to observable phenomena and metrics. This structure proved well-suited to the literature: each included article contributed to at least one of the three spreadsheets, with most included in two or more. These coding instruments, as well as more information on the process, are documented in full in the Supplementary Material (see Tables \ref{tab:S4_primary_coding}--\ref{tab:S6_metrics}).

Primary coding was performed by one researcher and two additional researchers were consulted for ambiguous cases. Specific coding fields were refined iteratively through initial coding of $\sim$10 articles and $\sim$3 of each subsequent new domain, with calibration discussions between three researchers to determine whether a new domain's methodological conventions required modifications to coding fields for consistency. These discussions established conventions for ambiguous cases: terms that exhibited inherently domain-specific differences were left with the original authors' terminology (e.g., cognitive "load" vs. "distraction"), whereas we kept consistent category terminology for domain-agnostic concepts such as metrics, induction names, and hazard names, for the purposes of statistical comparisons, referred to in Section \ref{sec:Results}. Formal inter-rater reliability was not computed; most coding fields captured factual study attributes with constrained response options, and consistency was maintained through the calibration conventions above and predefined dropdown menus.

\textbf{Supplementary Coding Tools.} While all data extraction and coding was completed manually using Google Sheets, the team supplemented with several other tools for the analysis process. AI tools Claude ~\cite{anthropic2025claude} and ChatGPT ~\cite{OpenAI2025ChatGPT} were used to supplement large-scale data comparisons, in particular for cross-referencing articles for unique article counts, checking coherency/consistency in claims, restructuring manuscript sections (including the abstract, introduction, and methods), and verifying citation formatting and accuracy. Derived statistics and AI-assisted contributions were independently verified by team members.

\section{Results}
\label{sec:Results}
The process outlined in the Methods section yielded a total of 91 papers across the 9 impairment domains (Table \ref{tab:Search Result Summary}). Of the 88 papers that evaluated study design, 75\% were simulator-only and 16\% were on-road only; 9\% included both simulator and on-road components. The majority of the articles focused on cognitive distraction and visual distraction, and the co-occurrence between these two domains (17 of 91) represented the highest rate across all of the domains.

\textbf{Notes on interpretation.} Two factors should inform how the following results are read. First, article counts vary substantially by domain; findings for stress, frustration, sudden sickness, readiness, and alcohol intoxication (each under 10 articles) should be viewed as preliminary. Second, as described in \ref{sec:Finding1}, simulator-only studies are not evenly distributed across domains. Domains that operate within an event-response paradigm rely on safety-critical events, which creates a structural requirement for simulated conditions. Performance-degradation oriented domains do not require a safety-critical event in order to assess continuous declines in metrics, thus are more feasibly studied on closed-road courses or public roads; however, a majority of these studies still opt for simulated designs for safety and ethical reasons. The reliance on simulated environments has real consequences for interpretation of study outcomes. For example, a validation study comparing intoxicated driving in a simulator against a closed test track found that lane-position variability (SDLP) increased with blood alcohol concentration in both settings, but the effect was larger in the simulator ~\cite{Helland20139}. This illustrates that even a metric with established validity can differ systematically in magnitude between simulated and real driving. Hazard integration is also weakest in performance-degradation domains; external validity concerns may compound the existing maturity gap in Fatigue and Intoxication studies.

\subsection{RQ1. How are induction methods standardized or varied within and across domains?}
\label{sec:RQ1}

Induction methods varied substantially in standardization across impairment domains. Two domains displayed clear convergence on dominant protocols: fatigue research consistently used a sleep restriction combined with caffeine control (9/17 induction instances); the second most common approach, time-of-day comparisons (4), manipulated the same underlying mechanism by scheduling drives in early morning or late night hours. Alcohol intoxication was consistently induced to manipulate BAC, although the specific modalities and BAC values varied. Readiness demonstrated the strongest protocol conformity: all 8 articles employed takeover tasks triggered by on-road events, though the alert modality (auditory, visual, haptic) varied across studies.

Inconsistency of induction design is apparent in the distraction, cognitive load, and stress domains. Cognitive distraction included 31 articles, but no dominant induction method: conversation tasks and N-back tasks each accounted for 7 instances, followed by texting while driving (4), phone tasks (4), in-vehicle office tasks (3), rotary mental visualization task (3), and over 20 other individual techniques appearing only once or twice. Visual distraction was similarly inconsistent, with 18 distinct tasks across 42 instances; arrow tasks (8), in-vehicle system interactions (6), and navigation tasks (5) formed loose clusters but none exceeded 20\% of instances. Cognitive load lacked both volume and convergence: only 3 of 10 articles studied cognitive load in isolation, and the tasks used overlapped heavily with cognitive distraction protocols. Stress methodologies consisted of 9 unique induction techniques across 11 instances, spanning physiological methods (e.g., electric shock, noise exposure), social threat paradigms (e.g., Trier Social Stress Test), and scenario-based manipulations (e.g., traffic density, time pressure). 

Two domains faced external constraints on induction feasibility. Sudden sickness (e.g., rapid onset of nausea, dizziness, and/or syncope) inductions were limited to blood-glucose manipulation (1) and simulator-induced motion sickness (2); the former is medically complex and the latter depends on hardware-specific parameters that limit replication. Frustration studies relied predominantly on scenario-based methods (e.g., aggressive drivers, traffic violations, slow lead vehicles) which are themselves hazard events. Only 3 of the frustration inductions used non-hazard methods: autobiographical recall (2) and pre-drive video clips (1).

\subsection{RQ2. Which hazard paradigms are used and how consistently within and across domains?}
\label{sec:RQ2}

Hazard integration patterns divided sharply across impairment domains. Distraction, readiness, and cognitive loading impairments demonstrated substantial hazard use: of 53 articles, 35 (66\%) included explicit reaction-inducing events. The most common event types were road obstructions (16), braking events (13), pedestrians (12), and scenario design elements (12). Hazards were used in 68\% of the reviewed cognitive distraction studies and 67\% of the reviewed visual distraction studies. Readiness studies integrate hazards by design: takeover tasks are inherently triggered by safety-critical events such as lane departures, lead vehicle braking, or system failures.
    
Frustration studies included the highest hazard density of any domain at 4.17 scenarios per article, with 22 hazard scenarios across 13 unique types. However, the function of these hazards differed from other domains: in 5 of 6 articles, hazards served to induce frustration rather than as critical events for measuring impaired response. Only one article used hazards to measure how a frustrated driver responds to danger ~\cite{Jeon2014_77}, which itself demonstrates the feasibility of such an approach.
    
Fatigue and alcohol intoxication studies reviewed included minimal hazard integration. Only three of 18 fatigue articles and three of eight intoxication articles included hazard events -- notably, the same three co-occurring articles. The remaining studies in these domains measured continuous metrics (e.g., lane deviation, speed variability, blink duration) without introducing discrete safety-critical events.
    
Reviewed stress articles included hazards in four of six articles, with all four instances (driver assistance failure, lead vehicle behavior, and skidding on ice) used for stress induction rather than as critical events for measuring impaired response. Cognitive load and sudden sickness sparsely included hazard integration with no consistent event type.

\subsection{RQ3. How do observable phenomena and metrics vary within and across domains?}
\label{sec:RQ3}
A total of 441 metric entries were coded across all impairments, spanning simulator telemetry, eye tracking, self-report instruments, physiological measures, camera and video observation, steering behavior, and reaction time (RT) measures.

Metric selection demonstrated domain-specific patterns. Fatigue and alcohol intoxication studies emphasized continuous measures: lane position \& speed variation, blink duration, EEG markers, and validated sleepiness scales (e.g., Karolinska Sleepiness Scale). Intoxication studies additionally anchored subjective measures against BAC as objective ground truth. Visual distraction studies converged on gaze-based metrics -- glance duration, eyes-off-road time, and gaze dispersion appeared consistently across the literature.

Cognitive distraction and readiness studies emphasized event-triggered measures -- RT, braking latency, time-to-collision (TTC), and takeover time. However, 17 of 31 cognitive distraction studies reviewed simultaneously investigated visual distraction, limiting clean attribution of metrics to either domain. Readiness metrics were internally consistent but all readiness studies reviewed co-induced an impairment to create conditions to compromise readiness, which prevents isolation of readiness-specific measurement patterns.

Subjective metric reliance varied substantially by domain (Table \ref{tab:rq3_subjective_ratios}). Frustration studies achieved the second highest subjective metric ratio at 32.5\% (13/40 metrics). Of these subjective measures, 92.3\% were emotion-confirmation instruments (e.g., Driving Anger Scale, Self-Assessment Manikin Categorical Scale) used to verify that the induction succeeded, rather than behavioral outcome measures. Alcohol intoxication study review resulted in a comparable subjective ratio (38.9\%), but these measures corroborated objective BAC data rather than serving as separate objective outcomes. Visual distraction had the lowest subjective ratio at 8.5\%.
\begin{table}[h]
    \centering
    \small
    \caption{Subjective Metric Ratios by Impairment Domain}
    \Description{Table ranking the nine impairment domains by the proportion of coded metrics that were subjective rather than objective, expressed as both a subjective-to-total count and a percentage. Alcohol Intoxication has the highest subjective ratio at 38.9 percent, followed by Frustration at 32.5 percent, Stress at 25.0 percent, and Fatigue at 24.8 percent. The remaining domains fall below 20 percent, with Visual Distraction at the lowest subjective ratio of 8.5 percent.}
    \label{tab:rq3_subjective_ratios}
    \begin{tabular}{@{}>{\centering\arraybackslash}p{0.4\linewidth}>{\centering\arraybackslash}p{0.25\linewidth}>{\centering\arraybackslash}p{0.25\linewidth}@{}}
        \toprule
        Domain & Subj./Total & Ratio \\
        \midrule
        Alcohol Intoxication  & 14/36  & 38.9\% \\
        Frustration           & 13/40  & 32.5\% \\
        Stress                & 7/28   & 25.0\% \\
        Fatigue               & 28/113 & 24.8\% \\
        Sudden Sickness       & 5/26   & 19.2\% \\
        Readiness             & 6/38   & 15.8\% \\
        Cognitive Distraction & 14/131 & 10.7\% \\
        Cognitive Load        & 5/52   & 9.6\%  \\
        Visual Distraction    & 12/142 & 8.5\%  \\
        \bottomrule
    \end{tabular}
\end{table}
\section{Discussion}

Our three RQs guided systematic comparison of induction methods, hazard paradigms, and observable phenomena and metrics across the nine impairment domains. As cross-domain patterns became apparent in the coded data, the variation across domains reflected systematic differences in methodological development and outcome interpretation that no existing impairment-specific framework was equipped to characterize. We therefore derived a maturity framework~\cite{keathley2016assessing} directly from the patterns our coding revealed, grounded in the same three dimensions of our RQs. These patterns also surfaced domain-specific anomalies that could not be captured by the spectrum alone; these became apparent in cases for which the structure and design of research within a domain, rather than its position on the spectrum, was the primary finding. It is important to note that this survey was designed to inform study design decisions rather than exhaustively review any single domain; maturity claims for high-coverage domains are made positively from the data, while for lower-coverage domains the claim is more limited.

\subsection{Finding 1: The Performance-Degradation vs. Event-Response Paradigm}
\label{sec:Finding1}
Impaired driving research has organized itself around two fundamentally different measurement paradigms -- per\-for\-mance-deg\-ra\-da\-tion and event-response -- without acknowledging that distinction. Rather than designing a study based on the nature of an impairment, the design is often inherited from convention.

Fatigue and intoxication research operates within a performance-degradation paradigm. These studies measure the degradation of driving performance or the erosion of vehicle control under impaired conditions, tracking metrics such as sustained lane-keeping quality, long-term speed consistency, or behavioral drift. As a result, these domains are studied as continuous states rather than as factors influencing driver response to discrete safety-critical events.

On the other hand, distraction and readiness research operates within an event-response paradigm. These studies employ scripted hazards that simulate safety-critical events, such as lead vehicle braking, pedestrian crossings, takeover tasks, and lane-changes, and measure the speed and accuracy of driver response under impaired conditions. Unlike degradation-focused studies, event-response studies focus on safety-critical event-response metrics, including RT, braking latency, TTC, and gaze behavior. However, these studies rarely focus on the continuous measures prioritized in performance-degradation studies. 

Cognitive Load is a unique case and was excluded from both paradigm clusters due to its ambiguous membership. Conceptually, sustained mental workload may be degradation; however, as discussed later, cognitive load induction methods overlap heavily with cognitive distraction, and its paradigm orientation remains undefined in the literature.

The isolation of impairment domains in either the performance-degradation or event-response paradigms is not necessarily inappropriate. It may reflect genuine differences in how impairments manifest; fatigue degrades performance and distraction primarily impairs discrete event responses. However, the field has not explicitly tested this; no study in this review directly compared both paradigms within the same participants under the same impairment conditions.

The implications for driver monitoring systems lie in optimization via fundamentally different signals with no guarantee that either generalizes across impairment domains. The variability in phenomena, instruments, and timescales limits the comparison of one domain to another; a common measurement language would make such comparisons valid. We return to this in Section~\ref{sec:Recommendations}.

The paradigm split also relates to the hazard integration pattern visible in Table  \ref{tab:maturity_dimension_ranking}. Performance-degradation paradigm domains lack hazard integration not because of methodological failure, but because the paradigm structurally excludes hazard events. Event-response paradigm domains inherently rely on hazard integration. Variations in impairment induction methodology and metric use across domains are independent of this paradigm split. The following section applies a maturity framework to each dimension, with paradigm membership informing the interpretation of hazard integration constraints.

\subsection{Finding 2: Maturity and Methodological Structure Across Impairment Domain Dimensions}

\textbf{Framework and Criteria.} For the purposes of this review, we assess research maturity along three dimensions that correspond to the RQs: induction consistency (RQ1), hazard integration (RQ2), and metric stability (RQ3). Each dimension is evaluated independently per impairment domain, using a three-level classification (with the latter two possibly overlapping):

\begin{enumerate}
\item \textbf{Mature:} The dimension demonstrates internal convergence. Dominant methods are identifiable, consistently applied acr\-oss studies, and reusable by other researchers.

\item \textbf{Fragmented:} Methods exist but lack convergence, either because no dominant approach has emerged or because the dimension cannot be cleanly isolated. The barriers are internal to the field.

\item \textbf{Constrained:} External barriers (ethics, safety, or feasibility) cap how rigorously the dimension can be studied. The lack of convergence reflects inherent boundaries on what can be done, not failure within the field.
\end{enumerate}

These three criteria were not imported from an existing framework, but emerged from patterns in the coded literature, and are proposed here as an analytical lens rather than a formal scoring instrument. We observed that dimensions fell short of maturity primarily in two ways, so we classified failure modes into `Fragmented' and `Constrained.' Both failure conditions prevent a domain from supporting cross-impairment comparison or generalizable driver monitoring systems, but they require distinctly different remedies: Fragmented dimensions need methodological standardization, while Constrained dimensions need technological or paradigmatic innovation to work around external limits. 

Notably, this framework decouples volume from maturity. A domain with many studies may still be Fragmented if those studies do not converge; a domain with few studies may still be Mature if the existing work shares a consistent methodology. For instance, fatigue research demonstrates convergence on sleep restriction as a dominant induction method despite appearing in only 18 articles, while cognitive distraction appeared in 31 articles with no single induction method accounting for more than 20\% of instances. Table \ref{tab:maturity_dimension_ranking} presents the dimension-level classifications for each impairment domain. The sections that follow provide the rationale for each classification.

\begin{table}
    \centering
    \scriptsize
    \setlength{\tabcolsep}{4pt}
    \caption{Maturity Dimension Ratings}
    \label{tab:maturity_dimension_ranking}
    \resizebox{\linewidth}{!}{%
    \begin{tabular}{@{}lcccccc}
        \toprule
         & \multicolumn{2}{c}{\textbf{Induction}}
         & \multicolumn{2}{c}{\textbf{Hazard}}
         & \multicolumn{2}{c}{\textbf{Metric}} \\
         & \multicolumn{2}{c}{\textbf{Consistency}}
         & \multicolumn{2}{c}{\textbf{Integration}}
         & \multicolumn{2}{c}{\textbf{Stability}} \\
         \cmidrule(lr){2-3}\cmidrule(lr){4-5}\cmidrule(lr){6-7}
         \textbf{Domain} & C & F & C & F & C & F \\
         \midrule
         Intoxication
          & \cellcolor{green!15}- & \cellcolor{green!15}-
          & \xmark & -
          & \cellcolor{green!15}- & \cellcolor{green!15}- \\
        Fatigue
          & \cellcolor{green!15}- & \cellcolor{green!15}-
          & \xmark & -
          & \cellcolor{green!15}- & \cellcolor{green!15}- \\
        Cognitive Distraction
          & - & \xmark
          & \cellcolor{green!15}- & \cellcolor{green!15}-
          & - & \xmark \\
        Visual Distraction
          & - & \xmark
          & \cellcolor{green!15}- & \cellcolor{green!15}-
          & \cellcolor{green!15}- & \cellcolor{green!15}- \\
         Stress*
           & \xmark & \xmark
           & - & \xmark
           & - & \xmark \\
         Frustration*
           & \xmark & \xmark
           & \cellcolor{green!15}- & \cellcolor{green!15}-
           & \xmark & \xmark \\
         Cognitive Load
           & - & \xmark
           & - & \xmark
           & - & \xmark \\
         Sudden Sickness
           & \xmark & \xmark
           & - & \xmark
           & \xmark & \xmark \\
         Readiness**
           & \cellcolor{green!15}- & \cellcolor{green!15}-
           & \cellcolor{green!15}- & \cellcolor{green!15}-
           & - & \xmark \\
       \bottomrule
   \end{tabular}
    }
    \vspace{1pt}
    \raggedright\scriptsize 
    Note: C = Constrained, F = Fragmented. Green shading indicates mature
    dimensions. *hazard used as impairment induction method; **takeover
    requires additional impairment use
    \Description{Table rating ten impaired-driving research domains on three methodological maturity indicators: Induction Consistency, Hazard
Integration, and Metric Stability. Each indicator has two sub-measures,
Constrained (C) and Fragmented (F). A green-shaded cell means the domain is
rated mature on that measure; an X mark means not mature; a dash means mixed or inconclusive evidence.

Intoxication: Induction Consistency mature on both Constrained and
Fragmented. Hazard Integration not mature on Constrained, mixed on
Fragmented. Metric Stability mature on both.

Fatigue: Induction Consistency mature on both. Hazard Integration not mature on Constrained, mixed on Fragmented. Metric Stability mature on both.

Cognitive Distraction: Induction Consistency mixed on Constrained, not
mature on Fragmented. Hazard Integration mature on both. Metric Stability
mixed on Constrained, not mature on Fragmented.

Visual Distraction: Induction Consistency mixed on Constrained, not mature
on Fragmented. Hazard Integration mature on both. Metric Stability mature on both.

Stress (marked as a domain where hazard exposure is used as the impairment
induction method): Induction Consistency not mature on both. Hazard
Integration mixed on Constrained, not mature on Fragmented. Metric Stability mixed on Constrained, not mature on Fragmented.

Frustration (marked as a domain where hazard exposure is used as the
impairment induction method): Induction Consistency not mature on both.
Hazard Integration mature on both. Metric Stability not mature on both.

Cognitive Load: Induction Consistency mixed on Constrained, not mature on
Fragmented. Hazard Integration mixed on Constrained, not mature on
Fragmented. Metric Stability mixed on Constrained, not mature on Fragmented.

Sudden Sickness: Induction Consistency not mature on both. Hazard
Integration mixed on Constrained, not mature on Fragmented. Metric Stability not mature on both.

Readiness (marked as requiring induction of an additional impairment, such
as through a takeover task, since it cannot be induced independently):
Induction Consistency mature on both. Hazard Integration mature on both.
Metric Stability mixed on Constrained, not mature on Fragmented.}
\end{table}

\textbf{Induction Consistency (RQ1)} Three domains meet the threshold for Mature induction consistency. Alcohol intoxication, fatigue, and readiness meet this threshold on the strength of the induction convergence already reported in Section \ref{sec:RQ1} (BAC modulation, sleep restriction, and universal takeover-task use, respectively). However, dangerous situations for drivers are ethically and practically constrained to the simulator, which is an increasingly common method of study for driver impairment. As a result, simulator validity concerns are not evenly distributed: they fall hardest on some of the domains with the highest real-world stakes.

Three domains are designated as having Fragmented induction consistency. Cognitive distraction and visual distraction lack an agreed-upon induction method despite their volume (Section \ref{sec:RQ1}). Cognitive load is Fragmented both by low isolation and by methodological overlap with cognitive distraction.

Three domains exhibit both Constraint and Fragmentation on induction; external barriers limit what can feasibly be studied, and the field has splintered in response. Sudden sickness faces ethical and practical barriers: blood-glucose manipulation is medically risky and sudden medical emergencies cannot be safely simulated. The field's workaround (simulator-induced motion sickness) is neither replicable across hardware setups nor representative of real-world incapacitation, while other techniques (e.g., blood-glucose manipulation) can be difficult to scale. This results in Fragmented methods that do not generalize to one another. Stress is Constrained by physiological heterogeneity: cortisol-mediated stress, sympathetic arousal, and psychological time pressure may be genuinely distinct phenomena requiring different induction approaches. Under this single label, the field has deployed many techniques, nine in this review, targeting different mechanisms without convergence. Frustration is Constrained by apparent dependence on hazardous scenarios for ecologically valid induction; the non-hazard alternatives that exist (autobiographical recall, pre-drive video) account for only three instances and have not coalesced into a viable second paradigm. In each case, the constraint came first, while the fragmentation is a downstream effect.

\textbf{Hazard Integration (RQ2)} Four domains meet the threshold for Mature hazard integration. Cognitive distraction and visual distraction include hazards in approximately two-thirds of articles (Section \ref{sec:RQ2}). Readiness integrates hazards by design: takeover tasks are inherently responses to safety-critical events. Frustration shows the highest hazard density of any domain, with hazard scenarios consistently present across studies, though these scenarios function as induction rather than dependent measures, creating a structural anomaly examined in Finding 3.

Three domains are classified as Fragmented. Cognitive load studies reviewed rarely used hazards. Stress includes hazards in most studies, but all instances function as stress induction rather than assessing impaired reactions. This is a parallel to the frustration inversion, though at lower volume and without the rich hazard infrastructure. Sudden sickness lacks consistent hazard events, likely limited upstream by induction constraints.

Two domains are Constrained on hazard integration. Fatigue and alcohol intoxication both had minimal hazard use (Section \ref{sec:RQ2}). In addition to study design constraints, which are described in Finding 3, there are ethical considerations that limit exposing intoxicated or fatigued participants to high-stress scenarios. As a consequence, the effect of these impairments in the highly dangerous moments that might cause a real-world accident remains largely uncharacterized.

\textbf{Metric Stability (RQ3).} Three domains meet the threshold for Mature metric stability. Alcohol intoxication benefits from BAC as objective ground truth, with behavioral and subjective metrics alike validated against it (Section \ref{sec:RQ3}). Fatigue research consistently reuses a rich set of established and cross-validated metrics across studies. Visual distraction converges on gaze-based metrics well-established in the eye-tracking literature, reflected in the lowest subjective ratio of any domain (Section \ref{sec:RQ3}).

Four domains are classified as having Fragmented metric stability. Cognitive distraction's metrics overlap so heavily with visual distraction (Section \ref{sec:RQ3}) that effects cannot be credited to one construct over the other. Cognitive load inherits this problem and lacks an objective anchor analogous to BAC, leaving subjective workload instruments as proxies without consistent validation. Stress research lacks a dominant measurement approach; physiological sensors are implemented, but without standardized protocols. Readiness metrics (takeover time, reaction time) are internally consistent but attribution to the effect of readiness cannot be isolated from the co-impairment required to induce the driver's unready state.

Two domains exhibit both constraint and fragmentation on metric stability. Sudden sickness is Constrained by the difficulty of safely and realistically inducing sudden medical events under research conditions. As described previously, the induction approaches that exist target fundamentally different phenomena, and this upstream fragmentation propagates to measurement: no shared metric framework has emerged because the studies are not measuring the same underlying condition. However, with only 5 articles in this review, this should be interpreted as early evidence of a pattern. Frustration metric stability is Constrained by the difficulty of inducing emotional states in a way that generalizes across individuals. Without an agreed-upon induction protocol analogous to BAC or sleep restriction, the field has defaulted in large part to scenario-based methods (aggressive drivers, traffic violations) that are ecologically valid but consume hazard events as stimuli. This creates downstream fragmentation in measurement: hazard response cannot be assessed as an outcome, leaving predominantly inward-pointing metrics. Frustration's subjective ratio (Section \ref{sec:RQ3}) reflects emotion-confirmation instruments rather than behavioral outcomes. The structural consequences of this inversion are examined further in Finding 3.

\subsection{Finding 3: Domain-Specific Anomalies and Their Implications for Unified Impairment Modeling}
\label{sec:Finding3}

The proposed maturity framework captures whether domains have converged methodologically, but some anomalies are structural rather than developmental. Two patterns emerged that the framework alone cannot address: conceptual ambiguity, in which domain boundaries are too blurred for isolated study, and structural misfits, in which study designs systematically circumvent the safety-critical questions that motivate the research.

\textbf{Categorical Misfits: Frustration and Fatigue.} Frustration presents a structural inversion. Hazard events function as the meth\-od to induce frustration rather than to test how impaired drivers respond, and outcomes are instead focused on whether the induction succeeded. This inversion is visible only in cross-domain comparison: within frustration-specific reviews, using hazards as the induction stimulus is a convention rather than an anomaly, since there is no adjacent domain against which the choice appears unusual. Hazard design as a dominant induction method also introduces issues with replication, limiting the generalizability of frustration-related findings and precluding meaningful comparison across studies that use different hazard designs. 

The consequences of the structural inversion propagate to measurement. The subjective measures heavily relied on in Frustration research point inward at the driver's emotional state, not outward at safety-focused outcomes. A rich subjective measurement set is not inherently problematic; however, they are treated as the primary claim with no anchoring to an objective metric. The field has invested in building the stimulus but not in studying the consequences, and safety-critical hazard responses under frustration conditions have therefore been neglected. This phenomenon is not limited to frustration. Stress demonstrates a parallel inversion, though at lower volume in our review: all 4 hazard instances in stress research function as stress induction rather than to allow for measures of impaired response. This convergence suggests that driver negative emotional state research may have a fundamental methodological inconsistency that warrants further examination.

Fatigue research operates within a performance-degradation paradigm that structurally excludes hazard events. This is not methodological oversight, but a practical constraint: introducing a safety-critical event risks alerting fatigued participants and disrupting the degradation that the study is designed to capture. However, the consequence is that fatigue's effect on discrete dangerous moments (the situations most likely to cause real-world crashes) remains largely unobserved. The domain is methodologically mature on induction and metrics, yet the central safety question (how does a fatigued driver respond when something goes wrong?) has not been systematically addressed. Both frustration and fatigue illustrate a broader pattern: a domain can appear well-studied by conventional assessment while leaving critical safety-relevant questions unanswered. 

\textbf{Structural Misfit: Readiness.} Readiness presents a different kind of structural problem: it cannot be studied as a standalone construct. Every readiness study in our review co-induced another impairment (typically a distraction) to create the conditions that compromise readiness. As a result, readiness has the strongest induction protocol convergence of any domain (all studies employed takeover tasks) paired with the weakest construct isolation (no study isolated readiness as the variable of interest). The metrics attributed to readiness (takeover time, reaction time, post-takeover lane stability) are inseparable from the co-induced impairment's effects on the same measures. This raises a question the review cannot resolve but should surface: whether readiness is appropriately classified as a driver impairment at all. The other domains in this review describe states in which the driver's capacity to drive is degraded by an internal or environmental factor. Readiness describes a transition failure; the driver is unable to resume control from an automated system at the moment of handoff. The state being measured is arguably not an impairment but a failure of the human-automation interface to support a timely transfer of control. We address this further in Section~\ref{sec:Recommendations}.

\textbf{Conceptual Ambiguity.} Cognitive distraction and cognitive load are rarely distinguished in study design despite their definitional differences in cognitive psychology. The implied distinction (load as passive workload, distraction as active secondary task engagement) does not hold consistently across the coded literature. Tasks classified as "cognitive load" (N-back, math problems, listening tasks) are operationally indistinguishable from those classified as "cognitive distraction," and only three of the ten cognitive load articles reviewed studied the construct in isolation.

Visual distraction faces a related problem. Naturalistic driving tasks tend to recruit both cognitive and visual attention simultaneously, and attempts to study one almost inevitably introduce the other. A navigation display task, for example, is primarily visual, but interpreting a route involves cognitive processing. As a result, 17 of 31 cognitive distraction articles simultaneously investigated visual distraction. In addition, visual distraction methodology overlaps with cognitive load and readiness.

These are not merely measurement problems. Cases for which conceptual clarity issues were relevant often were tied with inconsistent inductions, metrics, and hazard designs as a result. Conceptual clarity is a precondition for methodological maturity, not a parallel criterion. We propose concrete steps toward resolving this in Section~\ref{sec:Recommendations}.

\subsection{Recommendations}
\label{sec:Recommendations}

The following recommendations are preliminary, derived from a breadth-first audit rather than exhaustive domain coverage. Future cross-domain reviews and domain-specific empirical work should confirm or refine them.

\textbf{Resolve and define cognitive distraction and load.} We recommend the field adopt operational inclusion criteria that distinguish different cognitive task dimensions, such as passive workload manipulations (e.g., listening tasks with no response demand) from active secondary-task engagement (e.g., conversation or N-back tasks requiring ongoing response) or cognitive abilities activated, rather than relying on the task's label. Authors should also validate their choice of terminology against the foundational cognitive psychology literature rather than adopting a label by convention. This same scrutiny should extend to visual distraction tasks that recruit cognitive resources (e.g., a navigation display requiring route interpretation): such tasks should not be labeled purely cognitive distraction without accounting for their visual demands. Cognitive load research can then adopt an induction task that does not overlap with established cognitive distraction protocols, giving the domain a basis for independent methodological convergence.

\textbf{Decouple hazard-based induction from outcome measurement.} Frustration and stress require decoupling induction from outcome measurement (Finding 3). We recommend frustration studies adopt pre-drive induction methods (e.g., autobiographical recall, pre-drive video clips). For stress, we recommend induction protocols that target a specific mechanism (e.g., cortisol-mediated stress, time pressure) rather than treating ``stress'' as a single construct, as the apparent lack of convergence may partly reflect mixing stressor types together, rather than a true absence of standardization within any one type. For both domains, if a hazard-based induction is used, a clear separation between induction events and subsequent hazard events should be used to measure response, so that hazard events can serve their conventional role of eliciting and measuring a response under impaired conditions.

\textbf{Develop a bridging measurement paradigm.} The per\-for\-mance-deg\-rad\-ation and event-response paradigms (Finding 1) optimize for different signals with no established relationship between them. We recommend future studies adopt a composite design that measures both behavioral drift and event-response quality within a single impairment condition, which would allow direct empirical comparison of the two paradigms rather than continued reliance on convention. In the absence of a composite design, surrogates such as linkage to parallel existing studies, annotation of specific event-response intervals, or other alternatives are encouraged.

\textbf{Determine readiness's taxonomic status.} As discussed in Section \ref{sec:Finding3}, every readiness study in this review co-induced another impairment to compromise readiness, leaving the construct without a standalone measure. We recommend future work explicitly test whether readiness is better modeled as a driver impairment in its own right or as a downstream consequence of other impairments (e.g., distraction, fatigue) during automation handoff, since the current literature conflates the two. One such approach might involve comparing readiness-attributed metrics (e.g., takeover time, lane stability) across studies that use various co-induced impairments. If similar metrics are evaluated regardless of the impairment, that would suggest they capture something readiness-specific; if they instead vary with the particular impairment, that would support the view that what is being measured is actually indicative of a failure of the human-automation interface to support transfer of control.

\textbf{Establishing an agreed-upon refinement of the maturity framework.} In order to operationalize the maturity framework for the different domains, it would be beneficial for future research to standardize definitions of the taxonomy we used in a way that is agreed upon by the broader community. This could be accomplished via a Delphi process~\cite{nasa2021delphi} or similar approaches for consensus creation \cite{murphy1998consensus,fitch2001rand,schulz2010consort}, in order to distill and report on agreement or alternatives for guiding principles on maturity definitions.

\subsection{Limitations}

Despite our discussed implications for the field, the review does have several limitations. The breadth-first design meant uneven coverage of impairment domains, especially affecting the more studied domains, and preventing more deliberate use of the paper count as a consideration in maturity. For domains with fewer than 10 articles in this review (Stress, Frustration, Sudden Sickness, Readiness, Alcohol Intoxication), classifications should be interpreted as preliminary patterns rather than definitive assessments; absence of convergence in a small sample is consistent with fragmentation but does not prove it. We focused less on intoxication due to distinctions in study design for measurement and prevention, and because of our interest in using findings for study design purposes. Additionally, primary coding was performed by one researcher without formal inter-rater reliability assessment. Although calibration procedures and the largely deterministic nature of the coding fields mitigate this concern, future cross-domain reviews should incorporate multi-coder validation to strengthen confidence in domain classifications. While we include some literature that used on-road methodologies, our focus was on human-in-the-loop simulation studies. Finally, the maturity framework is proposed as an analytical lens, so domain placements are argued from the data instead of a formal rubric. 
\section{Conclusion}

In this paper we have surveyed human-in-the-loop simulator studies that probe driver impairment, focusing on 9 impairment domains. Our survey resulted in several key findings, including (1) identification of the performance-degradation vs. event-response paradigm split and its metric divergence, (2) a maturity framework for characterizing impaired driving research study dimensions, and (3) a cross-impairment comparison at sufficient breadth to reveal architectural patterns not visible in domain-specific reviews. 
We observe that impairment is widely studied but not characterized in a unified way, and would benefit from a shared methodological language across studies. Our findings raise the need to examine some impairments (e.g., frustration) in safety-critical situations, further standardize induction protocols (e.g., for cognitive load and stress) and measurement taxonomy, refine the differentiation of co-occurring phenomena (e.g., cognitive distraction, cognitive load, and visual distraction), and determine whether readiness is appropriately classified as a driver impairment at all.
Our review identifies several avenues for further exploration, such as studies that more consistently cover both behavioral drift and event-response aspects, or more standardized coverage of induction, scene, and metrics. Other lines of exploration involve the use of such unified studies to better inform and evaluate assistive ML approaches and a more robust characterization of these approaches across both simulator and on-road studies.

\begin{acks}
The authors thank Dr. Joshua Domeyer (Toyota Motor North America) for insightful discussions on multi-domain impaired driving research. The authors also thank Dr. Avinash Balachandran (Toyota Research Institute) for useful discussions and insights from the automotive perspective.
\end{acks}

\bibliographystyle{ACM-Reference-Format}
\bibliography{0_references.bib}


\clearpage
\appendix

\renewcommand{\thesection}{\arabic{section}}
\setcounter{section}{0}

\renewcommand{\thetable}{S\arabic{table}}
\renewcommand{\thefigure}{S\arabic{figure}}
\setcounter{table}{0}
\setcounter{figure}{0}

\section*{Supplementary Material}
\label{sec:supplementary_material}

\renewcommand{\thesection}{S\arabic{section}}
\setcounter{section}{0}
\section{Search Terms and Queries}

\subsection{Google Scholar Search Terms}

Table~\ref{tab:S1_google_scholar} lists all Google Scholar search terms that yielded results, along with the references for those results.

\subsection{Undermind Search Queries}

Table~\ref{tab:S2_undermind} lists all Undermind AI search queries that yielded results. Undermind search queries are created through a short back-and-forth discussion between the AI system and the user, which results in the search queries found below.

\section{Data Extraction Coding Schema}

This section documents the coding instruments used for data extraction and analysis. Six coding sheets were developed to systematically capture study characteristics, methodological dimensions, and screening decisions.

\textbf{Notes on Data Extraction:}
\begin{itemize}
    \item All coding was performed manually using Google Sheets, with AI tools (Claude, ChatGPT) used to supplement large-scale data comparisons and consistency checks as described in the Methods section.
    \item The three follow-up instruments (Induction Approaches, Incidents/Hazards, Observable Phenomena) correspond directly to the three research questions (RQ1--RQ3).
    \item Article identifiers (No.) link entries across all instruments to the Primary Coding Sheet.
\end{itemize}

\subsection{Primary Coding Sheet (Literature Survey)}

Table~\ref{tab:S4_primary_coding} documents the primary spreadsheet columns for extracting article-level data. Each included article received one row.

\subsection{Induction Approaches}

Table \ref{tab:S4_induction} catalogs impairment induction methods extracted from included articles.

\subsection{Incidents and Hazards}

Table \ref{tab:S5_hazards} catalogs hazard events and safety-critical scenarios extracted from included articles.

\subsection{Observable Phenomena and Metrics}
Table \ref{tab:S6_metrics} catalogs dependent variables and measurement instruments extracted from included articles.

\subsection{Process Documentation}
Table \ref{tab:S7_process} documents the search and screening process for reproducibility.

\subsection{Excluded Articles}
Table \ref{tab:S8_excluded} documents articles screened but excluded, with reasons, per PRISMA guidelines.

\subsection{Acronyms Reference}
Table \ref{tab:S9_acronyms} documents the columns for the spreadsheet with standardized abbreviations used across the coding instruments for quick reference.

\section{AI Tool Documentation (PRISMA-trAIce Compliance)}

\label{sec:prisma_traice}
This section documents the use of artificial intelligence (AI) tools in this systematic literature review, following the PRISMA-trAIce (Preferred Reporting Items for Systematic Reviews and Meta-Analyses---Transparent Reporting of Artificial Intelligence in Comprehensive Evidence Synthesis) checklist \cite{holst2025transparent}. Table~\ref{tab:s10_prisma_mapping} provides a complete mapping of all checklist items to their locations within this manuscript or supplementary material.

\subsection{Protocol and Registration (PRISMA-trAIce M1)}
\label{sec:s3_1_protocol}
No protocol was pre-registered for this review. The decision to use Undermind was made during the search phase after encountering limitations with keyword-based retrieval for certain impairment domains, particularly those with inconsistent terminology or requiring multi-concept queries (e.g., frustration, sudden sickness, readiness).

\subsection{AI Tool Identification (PRISMA-trAIce M2)}
Table~\ref{tab:s11_tool_identification} identifies the AI tools used, including version, developers, and each tool's role in the review process.

\subsection{Purpose and Stage of Application (PRISMA-trAIce M3)}
Table~\ref{tab:s12_purpose_stage} maps each tool to the systematic review stage in which it was applied.

\subsection{Input Data Description (PRISMA-trAIce M4)}
Table~\ref{tab:s13_input_data} describes the input data provided to each search method, including query format and development process.

\subsection{Output Data Description (PRISMA-trAIce M5)}
Table~\ref{tab:s14_output_data} describes the outputs returned by each search method and subsequent post-processing steps.

\subsection{Prompt Engineering (PRISMA-trAIce M6)}
\label{sec:s3_2_prompt}
Undermind queries are developed through iterative natural language dialogue rather than traditional LLM prompts. Users do not have access to model parameters (e.g., temperature, max tokens). Full query strings are documented in Table~\ref{tab:S2_undermind}.

\subsection{Operational Details and Settings (PRISMA-trAIce M7)}
Table~\ref{tab:s15_operational} describes Undermind's architecture, classification methodology, and reported performance benchmarks.

\subsection{Limitations of AI Tool Use (PRISMA-trAIce D1)}
\label{sec:S3.8_limitations_of_AI}
Table~\ref{tab:s16_limitations} summarizes known limitations of the AI tools used and the mitigation strategies employed.

\subsection{Human-AI Interaction and Oversight (PRISMA-trAIce M8)}
\label{sec:s3_3_human_ai}
A single reviewer processed title and full-text exclusion, evaluating Undermind search results based on the relevance of the title, Undermind summary, and relevance match score before proceeding to exclusion based on abstract. We note that relevant papers often scored $\sim$90\% and above on Undermind relevance match score, with exceptions down to ~$\sim$60\%. Some searches were iteratively refined based on initial results, with earlier iterations not fully reviewed if it was determined that the query was misaligned.

\subsection{Performance Evaluation (PRISMA-trAIce M9 and R2)}
\label{sec:s3_4_performance}
No independent performance evaluation was conducted for this review. Published benchmarks report approximately 98\% classification accuracy with no highly relevant articles misclassified as irrelevant \cite{hartke2024undermind}. See Table~\ref{tab:s15_operational} for additional performance metrics.

\subsection{Data Governance (PRISMA-trAIce M10)}
\label{sec:s3_5_governance}
The data source used was publicly available published literature only. There was no human subjects data collection, and no Institutional Review Board approval required. Extracted article metadata was stored in Google Sheets, and some full-text PDFs were stored on Google Drive as applicable for quick reference.

\subsection{Study Selection (PRISMA-trAIce R1)}
\label{sec:s3_6_selection}
Undermind was used solely for identification; all screening was performed manually. The PRISMA flow diagram (Figure 1) reflects this: details of the selection process outlined in Section \ref{sec:s3_3_human_ai} of the Supplementary Material. 

\subsection{Implications of AI Use (PRISMA-trAIce D2)}
\label{sec:s3_7_implications}
The use of Undermind provided several benefits for this review. The tool's iterative dialogue format enabled organic refinement of search parameters, with the system actively asking clarifying questions about scope. This is a more natural process than the repeated query reformulation and manual filtering required by traditional keyword search. This was particularly valuable given the review's breadth across nine impairment domains and its focus on study design characteristics rather than outcome results, which are difficult to capture in Boolean keyword syntax.

Several challenges were also encountered. Undermind operates as a black-box system; relevance percentage ratings were sometimes difficult to interpret, and the reasoning behind specific rankings was not transparent. Additionally, refining or re-running queries was time-consuming and resource-intensive, which limited iterative searches. Finally, Undermind's corpus is limited to arXiv, a preprint repository, which may not index all peer-reviewed venues relevant to driver impairment research. This limitation was mitigated through complementary Google Scholar searches.


\onecolumn
{\hbadness=10000
\begin{longtable}{l p{8.5cm} l p{7cm}}
\caption{Google Scholar Search Terms and Retrieved Articles}
\Description{Table listing nine Google Scholar queries with the number of articles retrieved by each and their reference numbers. Queries range from broad terms like "Driving and Impairment" to specific compound phrases. Article counts per query range from 1 to 17, with visual-distraction and cognitive-distraction queries yielding the most results (17 and 16 articles respectively). A note below the table indicates two additional articles were identified through citation chaining.}
\label{tab:S1_google_scholar}\\
\toprule
\# & Query String & Count & Refs \\
\midrule
\endfirsthead
\multicolumn{4}{l}{\small\textit{Table S1 continued from previous page}}\\[4pt]
\toprule
\# & Query String & Count & Refs \\
\midrule
\endhead
\midrule
\multicolumn{4}{r}{\small\textit{Continued on next page}}\\
\endfoot
\bottomrule
\endlastfoot

1  & Driving and ``visual distraction''
 & 17
 & \cite{Li2024_19,Rogers2011,Olsson2000,Shiferaw2014_32,Lee2010_33,Kaber2012_34,Amini2023_35,Yusoff2017_36,Muhrer2011_37,Kujal2016_38,Merat2016_39,Boyle2006_40,Lovell2022_41,Sun2023_42,Fancello2024_43,Chang2021_44,Ojsterek2023_45} \\[4pt]

2  & Driving and ``visual impairment''
& 1
& \cite{S2010_29} \\[4pt]

3  & Driving and Impairment
& 4
& \cite{Vacb2014_24,WOOD1994_26,Brookhuis2003_27,Fillmore2005_28} \\[4pt]

4  & Driving and cognitive AND distract
   & 16
   & \cite{Shi2012_51,Almahasneh2014_52,MedeirosWard2015_53,Farah2016,USA2013_55,Harbluk2007_56,Young2007,Liang2014_58,Kashevnik2021_59,Roberts2019_60,Sorum2022_61,Kuo2018_62,Marini2020_63,Wu2024_64,Xie2021_65,Zangi2022_66} \\[4pt]

5  & Driving and stress AND distraction
   & 6
   & \cite{Heimstra1970_67,Mathissen2021_68,Oliver2024_69,Deborne2008,Matthews1996_71,Dabic2023_72} \\[4pt]

6  & Frustration and ``driving performance''
   & 2
   & \cite{Xi2024_73,Maillant2025_74} \\[4pt]

7  & Angry and driving and performance
   & 1
   & \cite{Li2021_75} \\[4pt]

8  & Detecting and driver and ``sudden sickness''
   & 1
   & \cite{DeWilde2022_83} \\[4pt]

9  & ``cognitive load'' and driving and impairment and performance and simulator
   & 1
   & \cite{Felisberti2024_91} \\[4pt]

\end{longtable}
} 

Note: Articles \cite{Wang2024_76} and \cite{Jeon2014_77} were identified via citation chaining from queries 6 and 7 respectively, and are not counted as direct keyword retrievals above.

\clearpage
\twocolumn


{\hbadness=10000
\onecolumn
\begin{longtable}{l p{1.3cm} p{12cm} l p{1.5cm} }
\caption{Undermind Search Queries and Retrieved Articles}
\Description{Two-page table listing eleven Undermind AI natural-language search queries, organized by topic (cognitive load, sudden sickness detection, frustration, general impairment, readiness, and inducing sudden sickness). Each row includes the topic, the full natural-language query string used, the number of articles retrieved, and their reference numbers. Query strings are substantially longer than traditional keyword searches (typically several sentences), reflecting Undermind's dialogue-based query construction. Article counts per query range from 1 to 28, with the broadest impairment query yielding the highest count. A footnote defines SAE automation levels referenced in the queries.}
\label{tab:S2_undermind}\\
\toprule
\# & Topic & Query String & Count & Ref Nos. \\
\midrule
\endfirsthead
 
\multicolumn{5}{l}{\small\textit{Table S2 continued from previous page}}\\[4pt]
\toprule
\# & Topic & Query String & Count & Ref Nos. \\
\midrule
\endhead
 
\midrule
\multicolumn{5}{r}{\small\textit{Continued on next page}}\\
\endfoot
 
\bottomrule
\endlastfoot
 
1 & Cognitive load &
  I want to find driving simulator studies of licensed adult drivers that manipulate intrinsic cognitive load within the primary driving task (excluding cognitive distraction/secondary-task paradigms) across manual driving and automation supervision, report any cognitive-load measurements (including but not limited to standard deviation of lateral position (SDLP), subjective, physiological, and behavioral), and provide sufficient detail on study design and task manipulations such as roadway/environmental complexity, traffic density, hazard frequency, weather/visibility, route-guidance complexity, construction zones, and time pressure &
  3 & \cite{Kim2018_18,Li2022_90,Wang2024_92} \\[4pt]
 
2 & Cognitive load &
  I want to find driving simulator studies that manipulate or quantify intrinsic, extraneous, and/or germane cognitive load within the primary driving task and assess effects on driving performance, explicitly excluding secondary-task distraction paradigms &
  1 & \cite{Tao2025_49} \\[4pt]
 
3 & Detecting sudden sickness &
  I want to find empirical studies in fixed-base car driving simulators that detect acute medical phenomena (including simulator/cybersickness or motion sickness and other sudden-onset health events) using primarily driving-behavior signals (e.g., lane keeping, steering control, speed/pedal profiles), with physiological and self-report measures as secondary references, excluding fatigue and cognitive load &
  3 & \cite{Li2024_19,Xu2013_22,Smyth2018_81} \\[4pt]
 
4 & Frustration &
  I want to find experimental or quasi-experimental studies from the last 10--20 years that induce or verify driving-related frustration/irritation/anger and compare driving performance between frustrated and non-frustrated conditions (within- or between-subjects), conducted in simulators or on-road/closed-track settings, with licensed adult drivers or typical convenience samples, reporting at least one objective driving or safety metric (e.g., SDLP/lane deviation, speed or speed variability, brake/response time, headway/time to collision (TTC), collisions/near-misses), in human-driven or partially automated driving contexts &
  2 & \cite{Wang2023_78,Zhao2023_89} \\[4pt]
 
5 & Frustration &
  I want to find experimental studies from roughly the last 10 years (allowing earlier foundational work if needed) that explicitly induce driver frustration and compare driving performance within subjects between frustrated and non-frustrated conditions, including both simulator and on-road/closed-track settings, in representative samples of licensed adult drivers (excluding clinical or non-licensed groups), reporting objective outcomes with emphasis on SDLP/lane-keeping, speed and speed variability, and reaction/brake response time, while allowing other safety metrics (e.g., headway/time headway, hazard detection, collisions/near-misses), in human-driven and partially automated driving contexts &
  1 & \cite{Wang2023_78} \\[4pt]
 
6 & Frustration &
  I want to find experimental or quasi-experimental driving studies from the last 10--20 years that induce or verify frustration/irritation/anger and compare driving performance between frustrated and non-frustrated conditions (within- or between-subjects), conducted in simulators or on-road/closed-track settings, with licensed adult drivers or typical convenience samples, reporting at least one objective driving or safety metric (e.g., SDLP/lane deviation, speed or speed variability, brake/response time, headway/TTC, collisions/near-misses), in human-driven or partially automated (SAE L1--L2\textsuperscript{*}) driving contexts &
  1 & \cite{Wang2023_78} \\[4pt]
 
7 & Impairment &
  I want to find empirical studies (including systematic reviews) that quantitatively examine any type of driver impairment (such as distraction, fatigue, cognitive overload, or substance use) in driving contexts using simulator, on-road, or naturalistic methodologies, reporting on aspects of experimental design, impairment induction/measurement, environmental/task features, and, where relevant, the use of driver assistance or automation under impairment in all populations. &
  28 & \cite{Vakulin2007_1,Koopmans2023_2,Howard2007_3,Philip2005_4,Anund2008_5,Huizinga2019_6,Davies2001_7,Miller2019_8,Kang2017_9,Kundinger2020_10,Marando2022_11,Pachamuthu2024_12,RiminiDoering2005_13,Papantoniou2017_14,Dahlman2021_15,Lowrie2020_16,Guo2021_17,Kim2018_18,Li2024_19,Garca2013_20,soares2020drowsiness,Xu2013_22,Usman2024_23,Saito2016_25,Antony2023_47,TRK2019_48,Tao2025_49,Oguri2013_50} \\[4pt]
 
8 & Inducing sudden sickness &
  I want to find experimental or quasi-experimental driving studies from the last 10--20 years that induce sudden-onset acute sickness or medical-like impairment in adult licensed drivers---including but not limited to hypoglycemia, simulator/vection-induced motion or cybersickness, acute pain (e.g., cold pressor/pressure/capsaicin), presyncope proxies, mild normobaric hypoxia, migraine-like visual perturbations, vestibular perturbations, anxiogenic stressors, low-impact analogs (e.g., ``drunk goggles''), and feasible sedative-like manipulations---conducted in simulators or on-road/closed-track human-driven or SAE L1--L2 contexts (excluding L3--L5), that compare impaired versus non-impaired conditions (within or between-subjects), with any objective driving/safety metric when available, prioritizing in-session onset and using induction and/or physiological/symptom verification when available, while excluding studies focused solely on cognitive load or sleep deprivation unless explicitly tied to the induced sickness state &
  4 & \cite{Philip2005_4,Huizinga2019_6,Smyth2018_81,Hengstenberg2019_82} \\[4pt]
 
9 & Readiness &
  I want to find empirical human-subject simulator studies from 2014--present involving licensed adult drivers that examine attentional readiness (situational awareness, workload, vigilance) and secondarily physical readiness at the moment of automation-to-human handoff in Advanced Driver Assistance Systems (ADAS)/Automated Vehicle (AV) driving, emphasizing behavioral takeover performance and psychophysiology (including eye-tracking), covering both emergency and planned takeovers, and excluding reviews/modeling unless they are field-changing &
  6 & \cite{Gluck2022_84,Xu2024_85,Wu2024_86,Guo2023_87,Vogelpohl2019_88,Zhao2023_89} \\[4pt]
 
10 & Readiness &
  I want to find experimental or quasi-experimental driving studies from the last 10--20 years that assess automation takeovers/handoffs in SAE L2--L3 contexts, explicitly comparing planned versus unplanned handoffs and comparing ready versus non-ready driver conditions (within- or between-subjects), primarily in simulator settings (on-road/closed-track acceptable), with licensed adult or typical convenience samples, using readiness manipulations or classifications (e.g., type of non-driving related task (NDRT)/engagement, eyes-off-road state, alerting/time budget) and reporting objective takeover-quality and safety metrics (e.g., takeover request RT/latency, visual re-engagement, initial control input latency, maximum steering/deceleration rates, stability during the takeover window, post-takeover SDLP/lane position, speed/variability, headway/TTC, collisions/near-misses) &
  7 & \cite{Gluck2022_84,Xu2024_85,Wu2024_86,Guo2023_87,Vogelpohl2019_88,Zhao2023_89,Li2022_90} \\[4pt]
 
11 & Sudden sickness &
  I want to find experimental or quasi-experimental driving studies from the last 10--20 years that induce or verify sudden-onset acute medical impairment in adult licensed drivers (e.g., hypoglycemia, presyncope/syncope proxies, motion/cyber sickness, acute pain, hypoxia, migraine-like symptoms, seizure-like perturbations), comparing impaired versus non-impaired conditions (within- or between-subjects) in simulators or on-road/closed-track settings, in human-driven or partially automated (SAE L1--L2) contexts (excluding L3--L5), that report at least one objective driving/safety metric (e.g., SDLP/lane deviation, speed/variability, braking/response time, headway/TTC, collisions/near-misses), including induction and/or physiological verification when available; explicitly excluding purely cognitive-load or fatigue paradigms unless tied to a medically induced state or used as a comparison condition &
  1 & \cite{Stettler2023_80} \\
 
\end{longtable}
} 

\noindent{\footnotesize *SAE refers to Society of Automotive Engineers automation levels: L1 (driver assistance) and L2 (partial automation) require continuous driver supervision; L3 (conditional automation) and L4 (high automation) allow the system to handle driving tasks, with L3 requiring driver readiness to intervene and L4 operating without driver intervention within defined conditions.}
\twocolumn

\clearpage


\begin{table*}[!htbp]
\centering
\small
\renewcommand{\arraystretch}{1.25}
\caption{Summary Table of Impairment Article Counts and References}
\Description{Summary table showing the number of articles reviewed per impairment domain, broken into total articles (which may study the domain alongside others) and isolated articles (which study only that domain). Ten rows cover nine impairment domains plus an "Other" category. Total article counts range from 5 (sudden sickness) to 33 (visual distraction). The ratio of isolated to total articles varies substantially across domains: sudden sickness and frustration are studied entirely in isolation (5 of 5, 6 of 6), while readiness has zero isolated studies (0 of 8), reflecting its consistent co-induction with other impairments.}
\label{tab:search_result_summary}
\begin{tabular}{@{}>{\raggedright\arraybackslash}p{0.17\linewidth}>{\centering\arraybackslash}p{0.055\linewidth}>{\raggedright\arraybackslash}p{0.32\linewidth}>{\centering\arraybackslash}p{0.055\linewidth}>{\raggedright\arraybackslash}p{0.32\linewidth}@{}}
\toprule
Topic & Total & References & Isolated & References \\
\midrule
Cognitive Load & 10 & \cite{Marando2022_11,Kim2018_18,Li2024_19,Boyle2006_40,Tao2025_49,Xie2021_65,Gluck2022_84,Li2022_90,Felisberti2024_91,Wang2024_92} & 3 & \cite{Tao2025_49,Li2022_90,Felisberti2024_91} \\
Cognitive Distraction & 31 & \cite{Farah2016,USA2013_55,Harbluk2007_56,Young2007,Liang2014_58,Kashevnik2021_59,Roberts2019_60,Sorum2022_61,Kuo2018_62,Marini2020_63,Wu2024_64,Xie2021_65,Oliver2024_69,Wu2024_86,Guo2023_87,Vogelpohl2019_88,Zhao2023_89} & 11 & \cite{Guo2021_17,Oguri2013_50,Shi2012_51,Almahasneh2014_52,MedeirosWard2015_53,Farah2016,Harbluk2007_56,Young2007,Liang2014_58,Kuo2018_62,Marini2020_63} \\
Fatigue & 18 & \cite{Vakulin2007_1,Koopmans2023_2,Howard2007_3,Philip2005_4,Anund2008_5,Huizinga2019_6,Davies2001_7,Miller2019_8,Kang2017_9,Kundinger2020_10,Marando2022_11,RiminiDoering2005_13,Dahlman2021_15,Lowrie2020_16,soares2020drowsiness,Xu2013_22,Saito2016_25,Sorum2022_61} & 10 & \cite{Koopmans2023_2,Philip2005_4,Anund2008_5,Miller2019_8,Kang2017_9,RiminiDoering2005_13,Dahlman2021_15,soares2020drowsiness,Xu2013_22,Saito2016_25} \\
Readiness & 8 & \cite{Dabic2023_72,Gluck2022_84,Xu2024_85,Wu2024_86,Guo2023_87,Vogelpohl2019_88,Zhao2023_89,Wang2024_92} & 0 & \\
Alcohol Intoxication & 8 & \cite{Vakulin2007_1,Howard2007_3,Huizinga2019_6,Davies2001_7,Lowrie2020_16,Vacb2014_24,Fillmore2005_28,Shiferaw2014_32} & 3 & \cite{Vacb2014_24,Fillmore2005_28,Shiferaw2014_32} \\
Sudden Sickness & 5 & \cite{Lu2023,Stettler2023_80,Smyth2018_81,Hengstenberg2019_82,DeWilde2022_83} & 5 & \cite{Lu2023,Stettler2023_80,Smyth2018_81,Hengstenberg2019_82,DeWilde2022_83} \\
Stress & 6 & \cite{Heimstra1970_67,Mathissen2021_68,Oliver2024_69,Deborne2008,Matthews1996_71,Dabic2023_72} & 4 & \cite{Heimstra1970_67,Mathissen2021_68,Deborne2008,Matthews1996_71} \\
Frustration & 6 & \cite{Xi2024_73,Maillant2025_74,Li2021_75,Wang2024_76,Jeon2014_77,Wang2023_78} & 6 & \cite{Xi2024_73,Maillant2025_74,Li2021_75,Wang2024_76,Jeon2014_77,Wang2023_78} \\
Visual Distraction & 33 & \cite{Pachamuthu2024_12,Papantoniou2017_14,Kim2018_18,Li2024_19,Garca2013_20,Usman2024_23,Rogers2011,Olsson2000,Lee2010_33,Kaber2012_34,Amini2023_35,Yusoff2017_36,Muhrer2011_37,Kujal2016_38,Merat2016_39,Boyle2006_40,Sun2023_42,Fancello2024_43,Chang2021_44,Ojsterek2023_45,Antony2023_47,TRK2019_48,USA2013_55,Kashevnik2021_59,Roberts2019_60,Sorum2022_61,Wu2024_64,Zangi2022_66,Gluck2022_84,Xu2024_85,Guo2023_87,Vogelpohl2019_88,Zhao2023_89} & 10 & \cite{Pachamuthu2024_12,Usman2024_23,Olsson2000,Kujal2016_38,Sun2023_42,Fancello2024_43,Chang2021_44,Ojsterek2023_45,Antony2023_47,Zangi2022_66} \\
Other & 12 & \cite{Davies2001_7,Kundinger2020_10,Garca2013_20,WOOD1994_26,Brookhuis2003_27,S2010_29,Lovell2022_41,USA2013_55,Kashevnik2021_59,Wu2024_86,Guo2023_87,Vogelpohl2019_88} & 4 & \cite{WOOD1994_26,Brookhuis2003_27,S2010_29,Lovell2022_41} \\
\bottomrule
\end{tabular}
\end{table*}


\begin{table*}[htbp]
\centering
\caption{Primary Coding Sheet Fields}
\Description{Table listing the 23 field names and descriptions used in the primary coding spreadsheet, covering article metadata, sample characteristics, study format, impairment induction, scenario design, and results.}
\label{tab:S4_primary_coding}
{\small
\begin{tabular}{@{}>{\raggedright\arraybackslash}p{0.25\linewidth}>{\raggedright\arraybackslash}p{0.73\linewidth}@{}}
\toprule
\textbf{Field} & \textbf{Description} \\
\midrule
No. & Unique article identifier (integer) \\
Article Title & Full title of the publication \\
Year & Publication year \\
Access to article? & URL or access pathway \\
PDF on Drive & Boolean: whether full-text PDF was archived \\
Keywords & Search source (e.g., ``Undermind'', ``Google Scholar'') \\
Sample Size & Number of participants in the study \\
Goal/Relevant Info & Brief summary of study objectives \\
Study Format & Platform: ``Simulated'', ``On-Road'', or ``Simulated, On-Road'' \\
Measurements, Sensors, and Variables & Free-text field listing all dependent variables and instrumentation; prefixed with OBJ: (objective) or SUBJ: (subjective) \\
Driver assistance? & Boolean: whether ADAS or automation was present \\
Core Topic & Primary impairment domain(s) investigated \\
Impairment Induction or Task & Method used to induce the target impairment state \\
Interaction? & Boolean: whether the study examined interaction effects across impairments \\
Route Type & Driving environment (e.g., ``Highway'', ``Urban'', ``Rural'') \\
Route Features & Specific characteristics (e.g., ``traffic lights'', ``2-lane highway'') \\
Following Technique & Navigation method (e.g., ``Clear path'', ``Lead vehicle'') \\
Environmental Conditions & Time of day, weather, lighting \\
Traffic Density & Categorical: ``Light'', ``Moderate'', ``Heavy'' \\
Types of Traffic & Traffic composition (e.g., ``Vehicle'', ``Pedestrian'', ``Mixed'') \\
Reaction Inducement & Hazard events or safety-critical scenarios embedded in the study \\
Results & Key findings summary \\
Comments & Reviewer notes (e.g., sample restrictions, methodological notes) \\
\bottomrule
\end{tabular}
}
\end{table*}


\begin{table*}[htbp]
\centering
\caption{Induction Approaches Coding Fields}
\Description{Table listing the five field names and descriptions used in the induction approaches coding sheet: impairment domain, induction name, protocol description, article identifier, and source link.}
\label{tab:S4_induction}
\begin{tabular}{@{}>{\raggedright\arraybackslash}p{0.25\linewidth}>{\raggedright\arraybackslash}p{0.73\linewidth}@{}}
\toprule
\textbf{Field} & \textbf{Description} \\
\midrule
Impairment & Target impairment domain (e.g., ``Cognitive Distraction'', ``Fatigue'', ``Frustration'') \\
Induction Name & Short label for the induction method (e.g., ``N-back Task'', ``Sleep Restriction'') \\
Description/Instructions & Detailed protocol description as reported in the source article \\
No. & Reference to article identifier in Primary Coding Sheet \\
Link & URL to source article \\
\bottomrule
\end{tabular}
\end{table*}


\begin{table*}[htbp]
\centering
\caption{Incidents and Hazards Coding Fields}
\Description{Table listing the six field names and descriptions used in the hazards coding sheet: hazard name, impairment domain, scenario description, dependent measures, article identifier, and source link.}
\label{tab:S5_hazards}
\begin{tabular}{@{}>{\raggedright\arraybackslash}p{0.25\linewidth}>{\raggedright\arraybackslash}p{0.73\linewidth}@{}}
\toprule
\textbf{Field} & \textbf{Description} \\
\midrule
Incident/Hazard Name & Short label for the hazard type (e.g., ``Braking event'', ``Pedestrian crossing'') \\
Impairment & Impairment domain in which this hazard was used \\
Description & Detailed scenario description as reported in the source article \\
Stats & Dependent measures reported for this hazard (e.g., ``RT, collisions'') \\
No. & Reference to article identifier in Primary Coding Sheet \\
Link & URL to source article \\
\bottomrule
\end{tabular}
\end{table*}


\begin{table*}[htbp]
\centering
\caption{Observable Phenomena and Metrics Coding Fields}
\Description{Table listing the eight field names and descriptions used in the observable phenomena coding sheet, including subjective/objective classification, metric description, canonical category, required sensors, observability, impairment domain, article identifier number, and source link.}
\label{tab:S6_metrics}
\begin{tabular}{@{}>{\raggedright\arraybackslash}p{0.25\linewidth}>{\raggedright\arraybackslash}p{0.73\linewidth}@{}}
\toprule
\textbf{Field} & \textbf{Description} \\
\midrule
Subj. or Obj. & Classification: ``Subjective'' or ``Objective'' \\
Description & Detailed description of the metric as reported in the source article \\
Canonical Metric Category & Standardized metric label for cross-study comparison (e.g., ``SDLP'', ``RT'', ``Lane Position'') \\
Sensors/necessary tools & Instrumentation required (e.g., ``Simulator Telemetry'', ``Eye Tracker'', ``Survey'') \\
Observable? & Data collection method: ``Direct from Data'', ``Manual observation'', ``Not Observable'' \\
Impairment(s) & Impairment domain(s) in which this metric was used \\
No. & Reference to article identifier in Primary Coding Sheet \\
Link & URL to source article \\
\bottomrule
\end{tabular}
\end{table*}


\begin{table*}[htbp]
\centering
\caption{Process Documentation Coding Fields}
\Description{Table listing the seven field names and descriptions used to document the search and screening process: sequence order, topic, source platform, keywords, results URL, quality assessment, and notes.}
\label{tab:S7_process}
\begin{tabular}{@{}>{\raggedright\arraybackslash}p{0.25\linewidth}>{\raggedright\arraybackslash}p{0.73\linewidth}@{}}
\toprule
\textbf{Field} & \textbf{Description} \\
\midrule
Order & Sequence number of the search action \\
Topic & Target impairment domain for this search \\
Source & Search platform: ``Google Scholar'' or ``Undermind'' \\
Keywords/Search & Exact search query or keywords used \\
Results & URL to search results page \\
Perceived quality & Reviewer assessment of result relevance \\
Notes & Additional procedural notes (e.g., date restrictions, refinements) \\
\bottomrule
\end{tabular}
\end{table*}


\begin{table*}[htbp]
\centering
\caption{Excluded Articles Coding Fields}
\Description{Table listing the five field names and descriptions used to document excluded articles: article title, link, exclusion reason, date, and search source.}
\label{tab:S8_excluded}
\begin{tabular}{@{}>{\raggedright\arraybackslash}p{0.25\linewidth}>{\raggedright\arraybackslash}p{0.73\linewidth}@{}}
\toprule
\textbf{Field} & \textbf{Description} \\
\midrule
Article name & Full title of the excluded publication \\
Link & URL to the article \\
Reason & Exclusion rationale (e.g., ``Enough fatigue studies'', ``Not empirical'', ``Aviation context'') \\
Date Skipped & Date the exclusion decision was recorded \\
Source? & Search source from which the article was identified \\
\bottomrule
\end{tabular}
\end{table*}


\begin{table*}[htbp]
\centering
\caption{Acronyms Reference Fields}
\Description{Table listing the three field names used in the acronyms reference sheet: acronym, full expansion, and article identifiers where it appears.}
\label{tab:S9_acronyms}
\begin{tabular}{@{}>{\raggedright\arraybackslash}p{0.25\linewidth}>{\raggedright\arraybackslash}p{0.73\linewidth}@{}}
\toprule
\textbf{Field} & \textbf{Description} \\
\midrule
Acronym & Abbreviated term \\
Meaning & Full expansion \\
Where it appears & Article identifier(s) where the acronym is used \\
\bottomrule
\end{tabular}
\end{table*}


\begin{table*}[htbp]
\centering
\caption{PRISMA-trAIce Checklist Item Mapping}
\Description{Table mapping each of the items in the PRISMA-trAIce reporting checklist (organized as Title, Abstract, Introduction, Methods items M1–M10, Results items R1–R2, and Discussion items D1–D2) to their locations within the manuscript or supplementary material. Each row lists the item code, its description, and either the corresponding section or table reference, or a note that the item does not apply and why.}
\label{tab:s10_prisma_mapping}
\begin{tabular}{@{}p{1cm}p{4.5cm}p{11.5cm}@{}}
\toprule
\textbf{Item} & \textbf{Description} & \textbf{Location / Status} \\
\midrule
T1 & Title & Not applied; AI tools did not play a substantial role in the overall review process \\
A1 & Abstract & Not applied; same rationale as T1 \\
I1 & Introduction & Addressed in manuscript Methods (Section 3.1); placed in Methods rather than Introduction as search strategy context was more applicable \\
M1 & Protocol and Registration & Section~\ref{sec:s3_1_protocol} \\
M2 & Identification and Access & Table~\ref{tab:s11_tool_identification} \\
M3 & Purpose and Stage & Table~\ref{tab:s12_purpose_stage} \\
M4 & Input Data & Table~\ref{tab:s13_input_data} \\
M5 & Output Data & Table~\ref{tab:s14_output_data} \\
M6 & Prompt Engineering & Section~\ref{sec:s3_2_prompt}; limited applicability \\
M7 & Operational Details & Table~\ref{tab:s15_operational} \\
M8 & Human-AI Interaction & Section~\ref{sec:s3_3_human_ai} \\
M9 & Performance Evaluation (Methods) & Section~\ref{sec:s3_4_performance}; limited applicability \\
M10 & Data Governance & Section~\ref{sec:s3_5_governance} \\
R1 & Study Selection (AI-assisted) & Section~\ref{sec:s3_6_selection} \\
R2 & Performance Metrics (Results) & Section~\ref{sec:s3_4_performance}; limited applicability \\
D1 & Limitations of AI Use & Table~\ref{tab:s16_limitations} \\
D2 & Implications of AI Use & Section~\ref{sec:s3_7_implications} \\
\bottomrule
\end{tabular}
\end{table*}


\begin{table*}[htbp]
\centering
\caption{AI Tool Identification (PRISMA-trAIce Item M2)}
\Description{Comparison matrix identifying the three AI tools used in this review: Undermind AI, Claude, and ChatGPT. Rows describe eight attributes for each tool: tool name, version, developer, access URL, category, underlying technology, and role in the review. Undermind is identified as the primary literature-search tool built on GPT-4 with arXiv indexing; Claude (Opus 4.5) and ChatGPT (GPT-4) served supplementary analysis roles.}
\label{tab:s11_tool_identification}
\begin{tabular}{@{}p{3.5cm}p{5cm}p{4.65cm}p{3.5cm}@{}}
\toprule
\textbf{Attribute} & \textbf{Undermind AI} & \textbf{Claude} & \textbf{ChatGPT} \\
\midrule
Tool Name & Undermind & Claude & ChatGPT \\
Version & Cloud-based (not specified) & Claude 4.5 Opus (2026) & GPT-4 (2025) \\
Developer/Provider & Undermind, Inc. & Anthropic & OpenAI \\
Access URL & \url{https://www.undermind.ai} & \url{https://claude.ai} & \url{https://chat.openai.com} \\
Tool Category & AI-assisted semantic literature search & Large language model & Large language model \\
Underlying Technology & GPT-4 with arXiv \cite{arxiv} indexing & LLM (general) & LLM (general) \\
Role in Review & Literature search and retrieval & Supplementary analysis & Supplementary analysis \\
\bottomrule
\end{tabular}
\end{table*}


\begin{table*}[htbp]
\centering
\caption{Purpose and Stage of Application (PRISMA-trAIce Item M3)}
\Description{Table mapping each tool used to the systematic review stage in which it was applied and the specific task performed. Four rows cover: Undermind (identification stage, semantic literature discovery), Google Scholar (identification stage, keyword retrieval, non-AI), manual screening (no AI tool), and Claude/ChatGPT (data extraction stage, supplementary consistency checks alongside manual extraction).}
\label{tab:s12_purpose_stage}
\begin{tabular}{@{}p{3cm}p{3cm}p{11cm}@{}}
\toprule
\textbf{SLR Stage} & \textbf{Tool Used} & \textbf{Task Performed} \\
\midrule
Identification & Undermind AI & Semantic literature discovery using natural language queries \\
Identification & Google Scholar & Keyword-based literature retrieval (manual, non-AI) \\
Screening & N/A & Manual screening \\
Data Extraction & Claude / ChatGPT & Primarily manual extraction with large-scale checks \& cleanup suggestions by LLMs \\
\bottomrule
\end{tabular}
\end{table*}


\begin{table*}[htbp]
\centering
\caption{Input Data Description (PRISMA-trAIce Item M4)}
\Description{Comparison table describing the inputs provided to the two search methods used: Undermind AI and Google Scholar. Five attribute rows compare input type, input format, query development process, number of queries (11 Undermind vs. 9 Google Scholar), and topics covered by each method. Undermind was used for topics with inconsistent terminology or multi-concept requirements; Google Scholar covered topics where keyword syntax was effective.}
\label{tab:s13_input_data}
\begin{tabular}{@{}p{3cm}p{7cm}p{7cm}@{}}
\toprule
\textbf{Attribute} & \textbf{Undermind AI} & \textbf{Google Scholar (Non-AI)} \\
\midrule
Input Type & Natural language search queries & Boolean keyword queries \\
Input Format & Free-text descriptions of target study characteristics & Standard search syntax with AND operators and quoted phrases \\
Query Development & Iterative dialogue between user and AI system & Manual query formulation \\
Number of Queries & 11 & 9 \\
Topics Covered & Cognitive load, sudden sickness detection, frustration, general impairment, readiness/takeover & Visual distraction, cognitive distraction, stress, frustration, sudden sickness, cognitive load \\
\bottomrule
\end{tabular}
\end{table*}


\begin{table*}[htbp]
\centering
\caption{Output Data Description (PRISMA-trAIce Item M5)}
\Description{Comparison table describing the outputs returned by Undermind AI and Google Scholar. Four attribute rows cover output format, relevance scoring approach (Undermind uses proprietary semantic ranking; Google Scholar uses chronological or citation-based ordering), post-processing steps applied, and deduplication procedures.}
\label{tab:s14_output_data}
\begin{tabular}{@{}p{3.5cm}p{7.5cm}p{6cm}@{}}
\toprule
\textbf{Attribute} & \textbf{Undermind AI} & \textbf{Google Scholar} \\
\midrule
Output Format & Ranked list of article citations with bibliographic metadata in Undermind graphical user interface or exportable as PDF & Standard citation results \\
Relevance Scoring & Semantic relevance ranking (proprietary algorithm) & None (chronological or citation-based ordering) \\
Post-Processing & Results exported to spreadsheet; manual verification of relevance & Manual extraction of relevant articles; citation chaining \\
Deduplication & Cross-query deduplication performed manually & N/A \\
\bottomrule
\end{tabular}
\end{table*}


\begin{table*}[htbp]
\centering
\caption{Operational Details and Settings (PRISMA-trAIce Item M7)}
\Description{Table describing eight operational parameters of the Undermind AI tool, including its underlying language model (GPT-4), semantic search algorithm, arXiv corpus coverage (~2.3M papers), three-tier classification categories (highly relevant, closely related, ignorable), reported classification accuracy ($\sim$98\% with no highly-relevant-to-irrelevant misclassifications), query refinement mechanism, convergence behavior, and result limits. A notes column contextualizes each parameter.}
\label{tab:s15_operational}
\begin{tabular}{@{}p{3.5cm}p{7.5cm}p{6cm}@{}}
\toprule
\textbf{Parameter} & \textbf{Undermind AI} & \textbf{Notes} \\
\midrule
Underlying LLM & GPT-4 & Used for reasoning and classification \\
Algorithm Type & Semantic search with LLM-based relevance classification & Four-step process: basic search, classification, adaptation, convergence estimation \\
Coverage & arXiv corpus \cite{arxiv} ($\sim$2.3M papers) & Full-text search \\
Classification Categories & Highly relevant, Closely related, Ignorable & Assigned per paper based on query \\
Classification Accuracy & $\sim$98\% & No highly relevant $\rightarrow$ irrelevant misclassifications \cite{hartke2024undermind} \\
Query Refinement & AI-assisted iterative & User refines through dialogue \\
Convergence Behavior & $\sim$85\% of results found within first 150 papers evaluated & Exponential discovery curve \\
Result Limits & No hard limit; relevance-ranked until threshold not met & Max observed: 216 papers \\
\bottomrule
\end{tabular}
\end{table*}


\begin{table*}[htbp]
\centering
\caption{Limitations of AI Tool Use (PRISMA-trAIce Item D1)}
\Description{Table pairing four known limitations of the AI tools used in this review with their corresponding mitigation strategies. Limitations cover: arXiv-only corpus coverage (mitigated via complementary Google Scholar search), semantic matching gaps (mitigated via multiple query variations), reproducibility risks from cloud-based model updates (mitigated via full query documentation in Table S2), and opaque relevance ranking (mitigated via manual verification of all AI-retrieved results).}
\label{tab:s16_limitations}
\begin{tabular}{@{}p{0.5cm}p{8.25cm}p{8.25cm}@{}}
\toprule
\textbf{\#} & \textbf{Known Limitation} & \textbf{Mitigation Strategy} \\
\midrule
1 & \textbf{Coverage Uncertainty}: Undermind's corpus coverage is limited to arXiv; may not index all relevant databases or peer-reviewed venues & Complementary use of traditional keyword search (Google Scholar) to increase coverage \\
2 & \textbf{Semantic Matching Limitations}: Natural language queries may miss articles using different terminology & Multiple query variations for key topics (e.g., 3 frustration queries, 2 readiness queries) \\
3 & \textbf{Reproducibility}: Cloud-based tool with potential model updates; exact reproduction may not be possible & Full documentation of all query strings for transparency (see Table~\ref{tab:S2_undermind}) \\
4 & \textbf{Ranking Opacity}: Relevance scoring algorithm is proprietary & Manual verification of all AI-retrieved results \\
\bottomrule
\end{tabular}
\end{table*}
\end{document}